\documentclass{elsart}

\usepackage{amssymb}
\usepackage{graphicx}
\usepackage{color}

\definecolor{orange}{rgb}{1,0.5,0}

\begin{document}

\begin{frontmatter}



\title{Exact solutions for the two- and all-terminal reliabilities of the
Brecht-Colbourn ladder and the generalized fan}


\author{Christian Tanguy}

\address{France Telecom Division R\&D, CORE/MCN/OTT, 38--40 rue du G\'{e}n\'{e}ral Leclerc, 92794 Issy-les-Moulineaux Cedex, France}
\ead{christian.tanguy@orange-ft.com}


\begin{abstract}
The two- and all-terminal reliabilities of the Brecht-Colbourn ladder and
the generalized fan have been calculated exactly for arbitrary size as well as arbitrary individual edge and
node reliabilities, using transfer matrices of dimension four at most. While the all-terminal reliabilities of these graphs are identical, the special case of identical edge ($p$) and node ($\rho$) reliabilities shows that their two-terminal reliabilities are quite distinct, as demonstrated by their generating functions and the locations of the zeros of the reliability polynomials, which undergo structural transitions at $\rho = \displaystyle \frac{1}{2}$.
\end{abstract}

\end{frontmatter}

\section{Introduction}
\label{Introduction}




Network reliability has long been a practical issue since the pioneering work of Moore and
Shannon \cite{MooreShannon56}, and will remain addressed by reliability engineers, statistical physicists and
applied mathematicians for years, since networks are pervading our everyday life. Not surprisingly, the study of network reliability has led to a huge body of literature, which includes excellent textbooks and
surveys \cite{Ball95,Barlow65,Colbourn87,Shier91,Shooman68}. In the following, we consider a probabilistic approach of reliability, in
which the network is represented by an undirected graph $G = (V,E)$, where $V$ is a set of nodes
(also called vertices), and $E$ is a set of undirected edges (or links). Each element of $V$ and $E$ has a
probability $p_n$ or $p_e$ to operate correctly; failures of constituents are assumed
to occur at random, and to be statistically independent events (this restriction may be relaxed, eventually). Among the different measures of reliability, one often considers the $k$-terminal reliability, i.e., the probability that a given
subset $K$ of $k$ nodes ($K \subset E$) are connected. The most common instances are the
all-terminal reliability ${\rm Rel}_A$ ($K \equiv E$) and the two-terminal reliability ${\rm
Rel}_2(s \rightarrow t)$, which deals with a particular connection between a source $s$ and a
terminal destination $t$. Both of them are affine functions of {\em each} $p_n$ and $p_e$.

The sheer number of possible system states, namely $2^{|E|+|V|}$, clearly precludes the use of an
``enumeration of states'' strategy for realistic networks, and shows that the final expression may
be extremely cumbersome. Consequently, most studies have assumed graphs with perfect nodes ($p_n
\equiv 1$) and edges of identical reliability $p$; all reliabilities are then expressed as a polynomial in $p$, called the reliability polynomial. Radio broadcast networks  \cite{Graver05} have also been described by networks with perfectly reliable edges but imperfect nodes; in this context, one speaks of residual connectedness \cite{Boesch91}. It was shown early on \cite{Colbourn87} that the calculation of $k$-terminal reliability is \#P-hard, even after the simplifying and restricting assumptions that (i) the graph is planar (ii) all nodes are perfectly reliable (iii) all edges have the same reliability $p$.

The difficulty of the problem has stimulated many approaches : partitioning techniques \cite{Dotson79,Yoo88}, sum of disjoint products \cite{Abraham79,Balan03,Heidtmann89,Rai95,Soh93}, graph simplifications (series-parallel reductions \cite{MooreShannon56}, triangle-star (also called delta-wye) transformations \cite{Chari96,Egeland91,Gadani81,Rosenthal77,Wang96}, factoring \cite{KevinWood85}), determination of various lower and upper bounds to reliability polynomials \cite{Ball95,BrechtThesis85,BrechtColbourn86,BrownColbourn96,Chen04,Colbourn87,Colbourn91,Prekopa91,ScottProvan86}, Monte-Carlo simulations \cite{Fishman86,Karger01,Nel90}, genetic \cite{Coit96} and ordered binary decision diagram (OBDD) algorithms \cite{Kuo99,Rauzy03,Yeh02,Yeh02conf}. The reliability polynomial has also been studied \cite{Oxley02} with the aim of deriving some useful and hopefully general information from the structure of its coefficients \cite{Chari97,Colbourn93}, or the location of its zeros in the complex plane \cite{BrownColbourn92}.


The Tutte polynomial $T(G,x,y)$ of a graph $G$ has also been shown to be equivalent to the Potts model partition function
of the $q$-state Potts model \cite{Shrock00,Welsh00}. Calculations for various
recursive families of graphs $G$ quickly followed \cite{Chang03}. The all-terminal reliability polynomial ${\rm Rel}_A(G,p)$ of graphs is deduced from $T(G,1,\frac{1}{1-p})$. Royle and Sokal \cite{Royle04} proved that the Brown-Colbourn conjecture \cite{BrownColbourn92} on the location of the zeros of ${\rm Rel}_A(G,p)$, while valid for series-parallel reducible graphs, does not hold for a few families of graphs. While these results are extremely valuable to better understand a few properties of graphs and all-terminal reliability polynomials, they still assume that nodes are perfect and that edges have the same reliability.

In recent years, the growth of Internet traffic has called for a better evaluation
of the reliability of connections in --- among others, optical --- networks. This, of course, strongly depends on the
connection under consideration. Actual failure rates and maintenance data show that a proper
evaluation of two-terminal reliabilities must put node and edge equipments on an equal
footing, i.e., both edge (fiber links, optical amplifiers) and node (optical
cross-connects, routers) failures must be taken into account. The possibility of node failure has
been considered in early papers \cite{AboElFotoh89,Evans86,Hansler74}, to quote but a few.
Adaptation of algorithms to include imperfect nodes has been addressed, sometimes controversially \cite{Netes96,Ke97,Theologou91,Torrieri94,Yeh02conf}; the two-variable approach for
bounds to the reliability polynomial, by Bulka and Dugan \cite{Bulka94} and Chen and He
\cite{Chen04}, is also worth mentioning. In order to be realistic, different edge reliabilities
should be used: for instance, the failure rate of optical fiber links is likely to
increase with their length.

\vskip1cm
\begin{figure}[thb]
\hskip1cm
\includegraphics[width=0.75\linewidth]{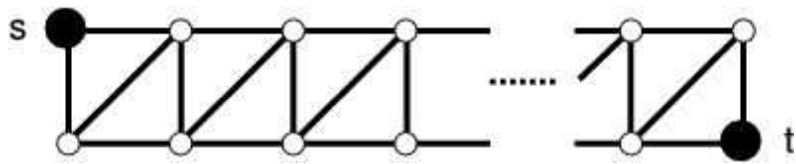}
\vskip0.3cm \caption{The Brecht-Colbourn ladder.}
\label{EchelleBC}
\end{figure}

\vskip1cm
\begin{figure}[thb]
\hskip3cm
\includegraphics[width=0.75\linewidth]{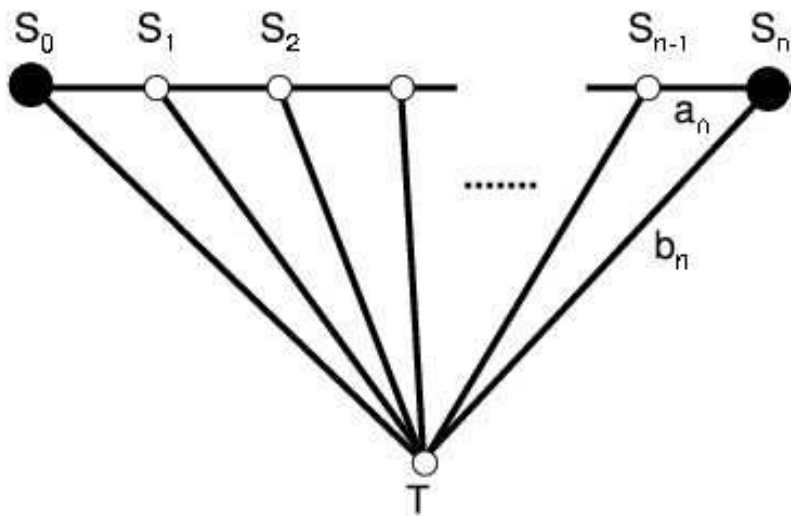}
\vskip0.3cm \caption{The generalized fan.} \label{Eventail}
\end{figure}
\vskip0.5cm


In this work, we give the exact two-terminal reliability of two well-known graphs, namely the Brecht-Colbourn ladder \cite{BrechtThesis85,BrechtColbourn86,Prekopa91} and the generalized fan \cite{Aggarwal75,Neufeld85}, which have the same all-terminal reliability. Both networks have been studied in the literature, especially the first one, as a case study for different lower bounds of the two-terminal reliability \cite{BrechtColbourn86}, but with a limited number of nodes. Here, the number of nodes will be arbitrary. Our solution fully exploits the recursive nature of the graphs and the triangle-star transformation with {\em unreliable} nodes \cite{Gadani81}. The final expressions are products of (at most) $4 \times 4$ transfer matrices, in which the arbitrary reliability of each node and edge appears explicitly. In the case of identical edge ($p$) and node ($\rho$) reliabilities, we provide the generating functions of the two-terminal reliabilities, leading to a very simple expression for the generalized fan. Another byproduct is the asymptotic locations of the zeros of the two-terminal reliability polynomials for the Brecht-Colbourn ladder and the generalized fan, which exhibit very different structures and undergo a structural transition at $\displaystyle \rho = \frac{1}{2}$.

Our aim is (i) to give a detailed derivation of the final results, so that people involved in reliability studies can readily use an easy-to-implement formula (ii) compare the exact solution with previous results (iii) exhibit the structural changes undergone by the location of the zeros of reliability polynomials as a function of node reliability (iv) emphasize anew the importance of algebraic structures of the underlying graphs in the determination of their associated polynomials \cite{Biggs93,Shier91}.

Our paper is organized as follows : In Section~\ref{Triangle-star
transformation}, we briefly recall the formulae for the
triangle-star transformation for unreliable nodes. In
Section~\ref{Main calculation}, we define the notations for the
different edge and node reliabilities and detail the derivation of
the main results (eqs.~(\ref{matricepassageBC}) and
(\ref{Rel2S0SnBC}) for the Brecht-Colbourn ladder,
eqs.~(\ref{MatriceTransfertEventail}) and (\ref{Rel2finalFan}) for
the generalized fan). Section~\ref{Identical reliabilities} is
devoted to the case where all edges and nodes have identical
reliabilities $p$ and $\rho$, respectively, the size of the network
appearing simply as an integer $n$. We give the analytical solution
of the two-terminal reliability ${\rm Rel}_2(p,\rho;n)$ for the
Brecht-Colbourn ladder and the generalized fan, and the associated
generating functions, which encode all the necessary information in
a beautifully simple, compact form. The asymptotic power-law or
constant behaviors are given when $n \rightarrow \infty$. Prompted
by the nearly universal character of the Brown-Colbourn conjecture
\cite{BrownColbourn92}, we address in Section~\ref{Zeros} the
location of zeros of the two-terminal reliability polynomials, and
show that their structures are very different even though their
internal structure is similar. For the sake of completeness, we
derive in Section~\ref{All-terminal reliability} the common
all-terminal reliability for both networks, for arbitrary values of
edge reliabilities. Finally, we conclude by indicating several
directions in which the present results may be further extended, so
that, for instance, a catalog of exactly solvable networks --- in
terms of reliability --- may be given rapidly \cite{CTpreparation}.
Such a catalog of elementary bricks could be useful for a new and
improved set of bounds or benchmarks for alternative methods in the
general case.

\section{Triangle-star transformation for unreliable nodes}
\label{Triangle-star transformation}

The triangle-star --- also called delta-wye or $\Delta-{\rm Y}$ --- transformation has been
used many times to simplify calculations of network reliability
\cite{Chari96,Colbourn87,Egeland91,Gadani81,KevinWood85,Rosenthal77,Wang96}. It has
mostly been applied in a perfect nodes context, to provide upper and lower bounds to the exact
reliability. Here, we exploit this transformation to the full in the case of imperfect nodes in
order to obtain exact results. Since it plays a crucial part of the derivation, we give the
formulae derived by Gadani \cite{Gadani81}.

\vskip0.5cm
\begin{figure}[htb]
\hskip1cm
\includegraphics[width=0.75\linewidth]{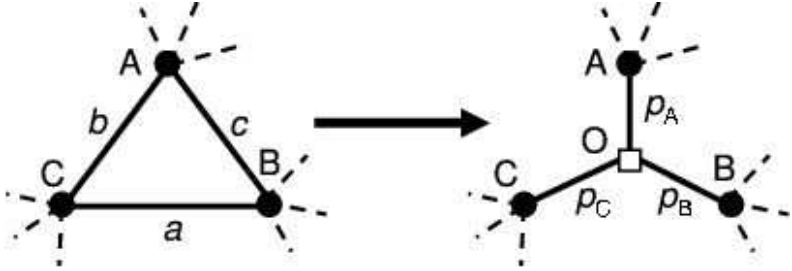}
\vskip0.8cm
\caption{Triangle-star transformation for unreliable nodes. $A$, $B$, and $C$ are the node reliabilities, with $a$, $b$, and $c$ the edge reliabilities of the initial network, and $p_A$, $p_B$, $p_C$, and $O$, those of the transformed network.}
\label{Delta-Y}
\end{figure}
\vskip0.5cm

Let us consider three particular nodes $A$, $B$, and $C$ of a network, connected as shown in the left part of Fig.~\ref{Delta-Y} (the reliability of the nodes are given by the same uppercase variables, so as to prevent an unnecessary multiplication of notations). The reliability of the edge connecting $A$ and $B$ is given by (lowercase) $c$, with similar notation for the remaining edges of the triangle. The aim of the triangle-star transformation is to replace the triangle by a star, which is possible by the addition of a new {\em unreliable} node $O$ and three new edges connecting $O$ to $A$, $B$, and $C$, with reliabilities $p_A$, $p_B$, and $p_C$, respectively. Both networks are equivalent provided that the following compatibility relations hold \cite{Gadani81}
\begin{eqnarray}
p_A \, O \, p_C & = & b + a \, c \, B - a \, b \, c \, B , \label{Delta-triangle AC}\\
p_A \, O \, p_B & = & c + a \, b \, C - a \, b \, c \, C , \label{Delta-triangle AB}\\
p_B \, O \, p_C & = & a + b \, c \, A - a \, b \, c \, A , \label{Delta-triangle BC}\\
p_A \, O \, p_B \, p_C & = & a \, b + b \, c + a \, c - 2 \, a \, b \, c . \label{Delta-triangle ABC}
\end{eqnarray}
Note that the first three equalities correspond to the probability
that the two nodes under consideration are connected, while the last
one gives the probability that the three nodes are connected. A word
of caution --- already given by Gadani --- is worth mentioning in
the case of {\em successive} triangle-star transformation: the
triangles should have no common edge or node.

\section{Derivation of the main results}
\label{Main calculation}

\subsection{Brecht-Colbourn ladder}

\begin{figure}[htb]
\hskip3.5cm
\includegraphics[width=0.75\linewidth]{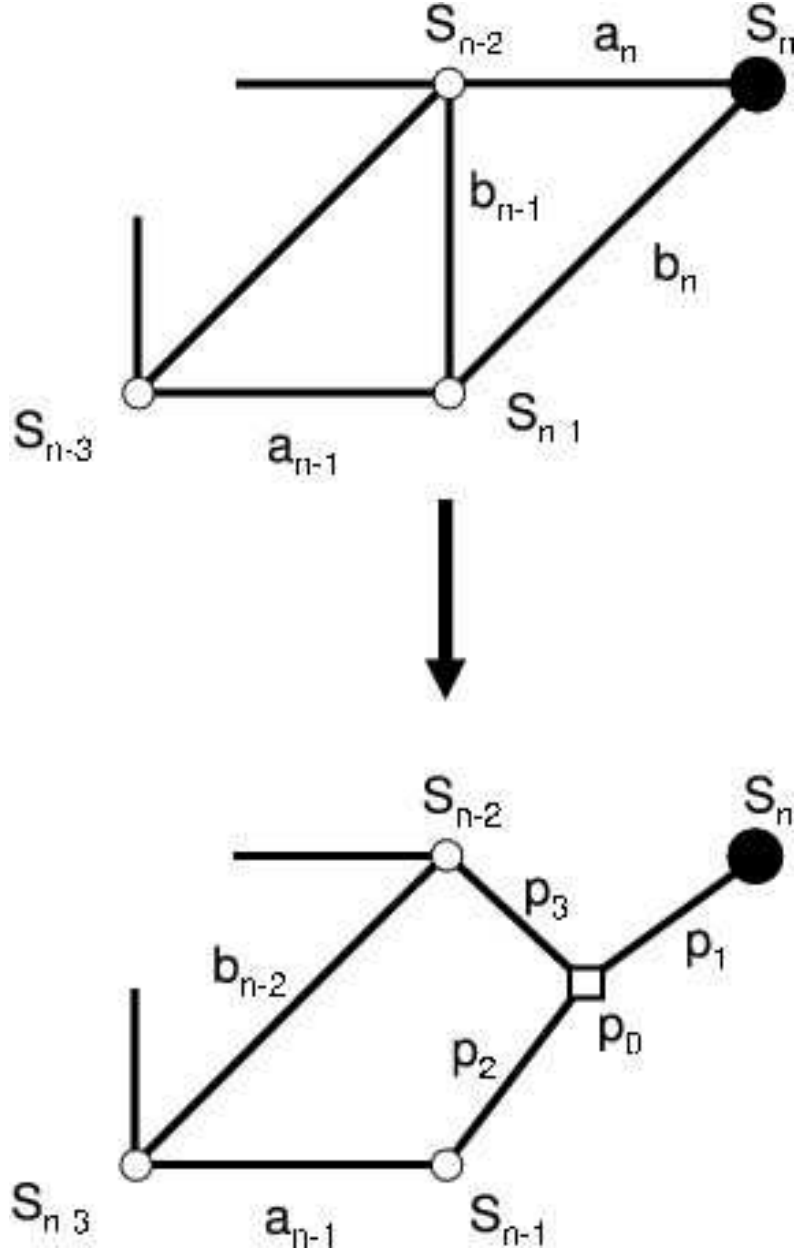}
\vskip0.3cm
\caption{Decomposition of the ladder using the triangle-star transformation.}
\label{Simplification Echelle BC}
\end{figure}

Let us first name the different edge and node reliabilities of the Brecht-Colbourn ladder, and detail how we can use the triangle-star transformation of the preceding section. Assuming that $S_0$ is always the source node, and using node reliabilities $S_i$, we note $a_n$ the reliability of the edge $(S_{n-2},S_{n})$, and $b_n$ that of $(S_{n-1},S_{n})$, as shown at the top of Fig.~\ref{Simplification Echelle BC}.

\vskip0.5cm
\vskip0.5cm

The application of the triangle-star transformation to Fig.~\ref{Simplification Echelle BC} leads to
\begin{eqnarray}
p_1 \, p_0 \, p_2 & = & b_n + a_n \, b_{n-1} \, S_{n-2} - a_n \, b_{n-1} \, b_{n} \, S_{n-2} , \label{transformeeBC1}\\
p_1 \, p_0 \, p_3 & = & a_n + b_{n-1} \, b_n \, S_{n-1} - a_n \, b_{n-1} \, b_n \, S_{n-1} , \label{transformeeBC2}\\
p_1 \, p_0 \, p_2 \, p_3 & = & a_n \, b_{n} + a_n \, b_{n-1} + b_{n-1} \, b_n - 2 \, a_n \, b_{n-1} \, b_n . \label{transformeeBC3}
\end{eqnarray}
Writing the two-terminal reliability ${\rm Rel}_2^{{\rm (BC)}}(S_0 \rightarrow S_n) = S_n \, {\mathcal S}_n$, we immediately see that
\begin{equation}
{\mathcal S}_n = p_1 \, p_0 \, {\mathcal S}_{n-1}(a_{n-1} \rightarrow a_{n-1} \, S_{n-1} \, p_2,
b_{n-1} \rightarrow p_3) , \label{recurrenceBCformelle}
\end{equation}
because the transformed graph remains essentially a Brecht-Colbourn ladder, provided that some edges are renormalized ($b_{n-1}$ must be replaced by $p_3$, $S_{n-1}$ by $p_0$, and $a_{n-1}$ by $a_{n-1} \, S_{n-1} \, p_2$).

In the case of imperfect nodes, the assumption
\begin{equation}
{\mathcal S}_n = \alpha_n \, a_n + \widetilde{\alpha}_n \, a_n \, S_{n-1} + \widetilde{\beta}_n \, b_n \, S_{n-1} + \widetilde{\gamma}_n \, a_n \, b_n \, S_{n-1}
\label{ansatzBCimperfectnodes}
\end{equation}
gives four parameters, for which we want to find a recursion relation. When we use eqs.~(\ref{transformeeBC1})--(\ref{transformeeBC3}) in eq.~(\ref{ansatzBCimperfectnodes}), there is one slightly tricky point: we are actually dealing with Boolean functions, so that $S_{n-2}^2 \equiv S_{n-2}$ (see for instance \cite{Aggarwal75}) and
\begin{eqnarray}
& & \hskip-2cm p_1 \, p_0 \, \widetilde{\alpha}_{n-1} \, S_{n-2} \, \left( a_{n-1} \, S_{n-1} \, p_2 \right) \nonumber \\
& = & \widetilde{\alpha}_{n-1} \, S_{n-1} \, S_{n-2} \, a_{n-1} \, (b_n
+ a_n \, b_{n-1} \, S_{n-2} - a_n \, b_{n-1} \, b_{n} \, S_{n-2})
\nonumber \\
& \equiv &  \widetilde{\alpha}_{n-1} \, S_{n-1} \, S_{n-2} \, a_{n-1}
\, (b_n + a_n \, b_{n-1} - a_n \, b_{n-1} \, b_{n}) .
\label{trickyBoolean1}
\end{eqnarray}
We deduce
\begin{equation}
\left(
\begin{array}{c}
\alpha_n \\
\widetilde{\alpha}_{n} \\
\widetilde{\beta}_{n} \\
\widetilde{\gamma}_{n}
\end{array}
\right)
 \;
 =
 \; M_{n-1}
 \; \cdot \; \left(
\begin{array}{c}
\alpha_{n-1} \\
\widetilde{\alpha}_{n-1} \\
\widetilde{\beta}_{n-1} \\
\widetilde{\gamma}_{n-1}
\end{array}
\right) ,
\label{factorisationavecmatricepassageBC}
\end{equation}
where the $4 \times 4$ transfer matrix $M_{n-1}$ is given by
\begin{equation}
M_{n-1} = \left(
\begin{array}{l|l|l|l}
0 & 0 & S_{n-2} & 0 \\ \hline
a_{n-1} \, b_{n-1} \, S_{n-2} & a_{n-1} \, b_{n-1} \, S_{n-2} & 0 & a_{n-1} \, b_{n-1} \, S_{n-2} \\ \hline
a_{n-1} & a_{n-1} \, S_{n-2} & b_{n-1} \, S_{n-2} & a_{n-1} \, b_{n-1} \, S_{n-2}\\ \hline
-a_{n-1} \, b_{n-1} \, S_{n-2} & - a_{n-1} \, b_{n-1} \, S_{n-2} &
- b_{n-1} \, S_{n-2} & a_{n-1} \, (1 - 2 \, b_{n-1}) \, S_{n-2}
\end{array}
\right) .
\label{matricepassageBC}
\end{equation}

For $n = 2$, the two-terminal reliability is easily solved because we have series-parallel graph:
\begin{equation}
{\rm Rel}_2^{{\rm (BC)}}(S_0 \rightarrow S_2) = S_2 \, (a_2 + b_1 \, S_1 \, b_2 - a_2 \, b_1 \, S_1 \, b_2) \, S_0 ,
\label{Rel2S0S2BC}
\end{equation}
which leads to $\alpha_2 = S_0$, $\widetilde{\alpha}_2 = 0$, $\widetilde{\beta}_2 = b_1 \, S_0$, and $\widetilde{\gamma}_2 = - b_1 \, S_0$, or equivalently to $\alpha_1 = \widetilde{\alpha}_1 = \widetilde{\gamma}_1 = 0$ et $\widetilde{\beta}_1 = 1$ (see eq.~(\ref{matricepassageBC})). The final result is
\begin{equation}
{\rm Rel}_2^{{\rm (BC)}}(S_0 \rightarrow S_n) = S_n \, (0 \; 0 \; 1 \; 0) \cdot M_n \, \cdot \, M_{n-1} \; \; \cdots \; \; M_1 \, \cdot \left(
\begin{array}{c}
0\\
0\\
1\\
0
\end{array}
\right) .
\label{Rel2S0SnBC}
\end{equation}
Note that in $M_1$, $a_1$ is quite arbitrary, so that it can be set equal to zero without loss of generality. Equation (\ref{Rel2S0SnBC}) also holds for $n$ equal to 1 or even 0 (following a frequent convention that a product of zero matrices is the identity matrix); it has also been independently checked for the first values of $n$, using a sum of disjoint products procedure.

In the case of perfect nodes, the ansatz should be ${\mathcal S}_n = \alpha_n \, a_n + \beta_n \, b_n + \gamma_n \, a_n \, b_n$. The relevant $3 \times 3$ transfer matrix ${M'}_n$ is now
\begin{equation}
{M'}_n = \left(
\begin{array}{c|c|c}
a_{n} \, b_{n} & 1 & a_{n} \, b_{n} \\ \hline
a_{n} & b_{n} & a_{n} \, b_{n} \\ \hline
- a_{n} \, b_{n} & - b_{n} & a_{n} \, (1 - 2 \, b_{n})
\end{array}
\right) ,
\label{matricepassageBCnoeudsparfaits}
\end{equation}
from which we deduce
\begin{equation}
{\rm Rel}_2^{{\rm (BC)}}(S_0 \rightarrow S_n;{\rm perfect \ nodes}) = (0 \; 1 \; 0) \cdot {M'}_n \, \cdot \, {M'}_{n-1} \; \; \cdots \; \; {M'}_1 \, \cdot \left(
\begin{array}{c}
0\\
1\\
0
\end{array}
\right) .
\label{Rel2S0SnBCnoeudsparfaits}
\end{equation}

\subsection{Generalized fan}

\vskip0.5cm
\begin{figure}[htb]
\hskip4cm
\includegraphics[width=0.75\linewidth]{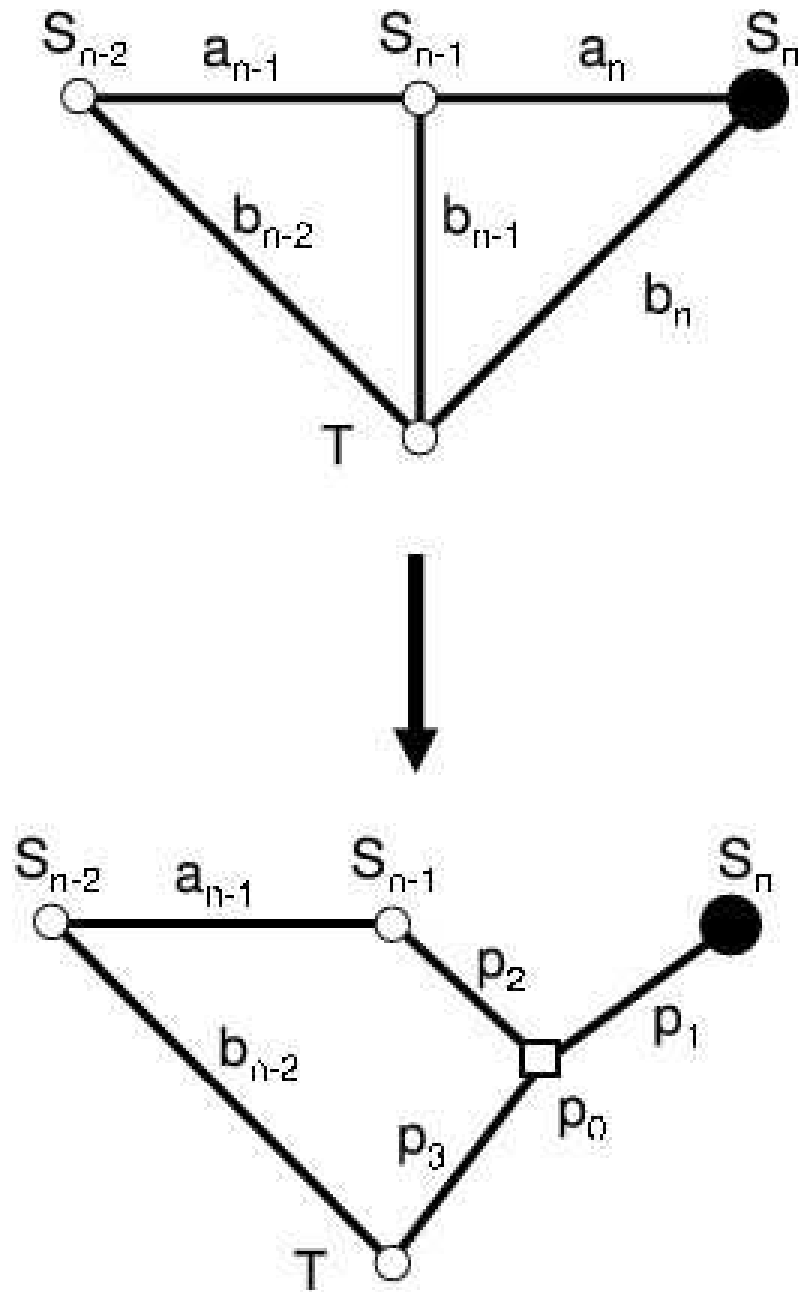}
\vskip0.3cm
\caption{Triangle-star transformation for the generalized fan.}
\label{Delta-Y pour Fan}
\end{figure}
\vskip0.5cm

Let us now turn to the second architecture. The triangle-star transformation, applied to the generalized fan, is displayed on Fig.~\ref{Delta-Y pour Fan}, and we expect the common node $T$ to play a special role in the recursion relation. We can write ${\rm Rel}_2^{{\rm (fan)}}(S_0 \rightarrow S_n) = S_n \; \widehat{{\mathcal S}}_n$, and see immediately that
\begin{equation}
\widehat{{\mathcal S}}_n = p_1 \, p_0 \, \widehat{{\mathcal S}}_{n-1}(a_{n-1} \rightarrow a_{n-1} \, S_{n-1} \, p_2,
b_{n-1} \rightarrow p_3) ,
\label{recurrenceBCformelle}
\end{equation}
with
\begin{eqnarray}
p_1 \, p_0 \, p_2 & = & a_n + b_{n-1} \, b_n \, T - a_n \, b_{n-1} \, b_n \, T , \label{transformeeFan1}\\
p_1 \, p_0 \, p_3 & = & b_n + a_n \, b_{n-1} \, S_{n-1} - a_n \, b_{n-1} \, b_{n} \, S_{n-1} , \label{transformeeFan2}\\
p_1 \, p_0 \, p_2 \, p_3 & = & a_n \, b_{n} + a_n \, b_{n-1} + b_{n-1} \, b_n - 2 \, a_n \, b_{n-1} \, b_n .\label{transformeeFan3}
\end{eqnarray}
Our ansatz is now
\begin{equation}
\widehat{{\mathcal S}}_n = S_n \, \alpha_n \, a_n + \widetilde{\alpha}_n \, a_n \, T + \widetilde{\beta}_n \, b_n \, T + \widetilde{\gamma}_n \, a_n \, b_n \, T .
\label{ansatzFanimperfectnodes}
\end{equation}
Here again, we must be careful with the Boolean variable $T$. When using eq.~(\ref{transformeeFan1}) in eq.~(\ref{ansatzFanimperfectnodes}), all occurrences of $T^2$ must be replaced by $T$. The recursion relations take the now familiar form
\begin{equation}
\left(
\begin{array}{c}
\alpha_n\\
\widetilde{\alpha}_n\\
\widetilde{\beta}_n\\
\widetilde{\gamma}_n
\end{array}
\right)
=
\widehat{M}_{n-1}
\cdot
\left(
\begin{array}{c}
\alpha_{n-1}\\
\widetilde{\alpha}_{n-1}\\
\widetilde{\beta}_{n-1}\\
\widetilde{\gamma}_{n-1}
\end{array}
\right) ,
\label{RecurrenceFan}
\end{equation}
where the transfer matrix $\widehat{M}_{n}$ is
\begin{equation}
\widehat{M}_n =
\left(
\begin{array}{c|c|c|c}
a_n \, S_n & 0 & 0 & 0 \\ \hline
0 & a_n \, S_n & b_n \, S_n & a_n \, b_n \, S_n \\ \hline
a_n \, b_n \, S_n & a_n \, b_n \, S_n & 1 & a_n \, b_n \, S_n\\ \hline
- a_n \, b_n \, S_n & - a_n \, b_n \, S_n  & - b_n \, S_n  & a_n \, (1 - 2 \, b_n) \, S_n
\end{array}
\right) .
\label{MatriceTransfertEventail}
\end{equation}
When $n=1$, we have a series-parallel graph:
\begin{equation}
{\rm Rel}_2^{{\rm (fan)}}(S_0 \rightarrow S_1) = S_1 (a_1 + b_0 \, b_1 \, T - a_1 \, b_0 \, b_1 \, T) \, S_0 ,
\label{Rel2S0S1Fan}
\end{equation}
from which we deduce the final expression
\begin{equation}
{\rm Rel}_2^{{\rm (fan)}}(S_0 \rightarrow S_n) = (1 \; T \; 0 \; 0) \, \cdot \, \widehat{M}_n \, \cdot \widehat{M}_{n-1} \; \cdots \; \widehat{M}_1 \, \cdot \, \widehat{M}_0 \, \cdot \,
\left(
\begin{array}{c}
1 \\
0 \\
0 \\
0
\end{array}
\right) ,
\label{Rel2finalFan}
\end{equation}
with the additional convention $a_0 \equiv 1$ for $\widehat{M}_0$. We have also checked the correctness of eq.~(\ref{Rel2finalFan}) by a sum of disjoint products procedure for the first values of $n$. It also agrees with the expression of Aggarwal {\em et al.} \cite{Aggarwal75} for $n=3$.

In the case of perfect nodes, which is often considered in the literature, the ansatz $\widehat{{\mathcal S}}_n = \alpha_n \, a_n + \beta_n \, b_n + \gamma_n \, a_n \, b_n$ leads to
\begin{equation}
{\rm Rel}_2^{{\rm (fan)}}(S_0 \rightarrow S_n; {\rm perfect \ nodes}) = (1 \; 0 \; 0) \, \cdot \, \widehat{M'}_n \, \cdot \widehat{M'}_{n-1} \; \cdots \; \widehat{M'}_1 \, \cdot \, \widehat{M'}_0 \, \cdot \,
\left(
\begin{array}{c}
1 \\
0 \\
0
\end{array}
\right) ,
\label{Rel2S0Snnoeudsparfaits}
\end{equation}
with
\begin{equation}
\widehat{M'}_n =
\left(
\begin{array}{c|c|c}
a_n & b_n & a_n \, b_n \\ \hline
a_n \, b_n & 1 & a_n \, b_n \\ \hline
- a_n \, b_n & - b_n & a_n \, (1 - 2 \, b_n)
\end{array}
\right) ,
\label{MatriceTransferEventailNoeudsParfaits}
\end{equation}
and again $a_0 \equiv 1$ for $\widehat{M}_0$.

It is worth noting that while the sizes of $M_n$ and $\widehat{M}_n$ for the Brecht-Colbourn and fan cases are identical, their matrix elements are clearly distinct. This actually leads to strong differences, as will be shown in the following section.

\section{Identical reliabilities $p$ and $\rho$}
\label{Identical reliabilities}

\subsection{Introduction}

While eqs.~(\ref{matricepassageBC}) and (\ref{Rel2S0SnBC}) give the two-terminal reliability for the
general Brecht-Colbourn ladder, and eqs.~(\ref{MatriceTransfertEventail}) and (\ref{Rel2finalFan}) for the fan, we consider here the special case where all edges and nodes have reliabilities $p$ and $\rho$, respectively. Because identical edge reliabilities are usually taken for granted, our exact results may help to demonstrate once again the underlying connection between combinatorics and reliability theory, most particularly in the enumeration of self-avoiding walks \cite{Colbourn87,Graver05,Prekopa91}. This assumption greatly simplifies the problem, since all transfer matrices, with the exception of $\widehat{M}_0$ in the case of the generalized fan, are identical to a matrix $M(p,\rho)$ (or $\widehat{M}(p,\rho)$): the two-terminal reliability is essentially the $n^{\rm th}$ power of one matrix. While the determination of its eigenvalues and eigenvectors is one way to solve the problem, we adopt in the following the generating function formalism, which is a fundamental tool in combinatorics \cite{Stanley97}, and makes for a very compact synthesis of the results. The generating function $G(z)$ is defined by
\begin{equation}
G(z) = \sum_{n=0}^{\infty} {\rm Rel}_2^{(n)}(p,\rho)  \, z^n .
\label{developpement
general de fonction generatrice}
\end{equation}
From the finite-order recursion relations obeyed by ${\rm Rel}_2^{(n)}$, the generating function is obviously a rational function of $z$, namely $G(z) = N(z)/D(z)$. In the present study, its denominator is at most of degree 4 in $z$, because in the absence of further simplification, $D(z) = z^4 \, P_{\rm carac}(1/z)$, where $P_{\rm carac}$ is the characteristic polynomial of the transfer matrix. The determination of $N(z)$ is then very straightforward: we only need to multiply the first terms of the expansion of $G(z)$ by $D(z)$ to see the numerator $N(z)$ emerge. The partial fraction decomposition of $G(z)$ finally leads to a compact, sometimes very simple, analytical expression of ${\rm Rel}_2^{(n)}(p,\rho)$, for arbitrary $n$.

\subsection{Brecht-Colbourn ladder}

\subsubsection{Imperfect nodes and edges}

When edge and node reliabilities are $p$ et $\rho$, the transfer matrix of eq.~(\ref{matricepassageBC}) is equal to
\begin{equation}
M(p,\rho) =
\left(
\begin{array}{l|l|l|l}
0 & 0 & \rho & 0 \\ \hline
p^2 \, \rho & p^2 \, \rho & 0 & p^2 \, \rho \\ \hline
p & p \, \rho & p \, \rho & p^2 \, \rho\\ \hline
-p^2 \, \rho & - p^2 \, \rho & - p \, \rho & p \, (1 - 2 \, p) \,
\rho
\end{array}
\right) ,
\label{M(p,rho) BC noeuds imparfaits}
\end{equation}
and
\begin{equation}
{\rm Rel}_2(S_0 \rightarrow S_n) = \rho \, (0 \; 0 \; 1 \; 0) \cdot M(p,\rho)^n \, \cdot
\left(
\begin{array}{c}
0\\
0\\
1\\
0
\end{array}
\right) .
\label{Rel2S0Sn BC p rho}
\end{equation}
The characteristic polynomial of this matrix is
\begin{eqnarray}
P_{\rm carac}^{\rm (BC)}(p,\rho;x) & = & x^4 - p \, (2 - p) \, \rho \, x^3 - p \, \rho \, \big(1- \rho \, (p+p^2-p^3)\big) \, x^2 \nonumber \\
& & + (1-p) \, (1- p \, \rho) \, p^2 \, \rho^2 \, x - (1-p) \, (1- \rho) \, p^4 \, \rho^3 .
\label{P caracteristique BC noeuds imparfaits}
\end{eqnarray}
The eigenvalues of $M(p,\rho)$ are given by the roots of this polynomial of degree 4. From the first few values of ${\rm Rel}_2(S_0 \rightarrow S_n)$ (taking $n = 2, ..., 10$, for instance), we can easily deduce the generating function $G_{\rm BC}^{(0)}(z)$.
\begin{equation}
G_{\rm BC}^{(0)}(z) = p \, \rho^2 \, z^2 \; \frac{N_{\rm BC}^{(0)}(z)}{D_{\rm BC}(z)} ,
\label{G BC p rho}
\end{equation}
with
\begin{eqnarray}
N_{\rm BC}^{(0)}(z) & = & 1 + p \, (1-p) \, \rho + p^2 \, \rho (1 - p \, \rho \, (2 - p)) \, z - p^2 \, \rho^2 \, (1-p)^2 \, z^2 \nonumber \\
& & + p^4 \, (1-p) \, \rho^3 \, (1-\rho) \, z^3 \label{N BC p rho BIS} \\
D_{\rm BC}(z) & = & 1 - p \, (2 - p) \, \rho \, z - p \, \rho \, \big(1- \rho \, (p+p^2-p^3)\big) \, z^2 \nonumber \\
& & + (1-p) \, (1- p \, \rho) \, p^2 \, \rho^2 \, z^3 - (1-p) \, (1- \rho) \, p^4 \, \rho^3 \, z^4
\label{D BC p rho BIS}
\end{eqnarray}
Adding $\rho + p \, \rho^2 \, z$ (basically, the two-terminal reliability of a single node, or that of two connected nodes) to $G_{\rm BC}^{(0)}(z)$, the numerator further simplifies, so that we may use another generating function, which of course provides the same coefficient of $z^n$ for $n \geq 2$, namely
\begin{equation}
G_{\rm BC}(z) = \rho \; \frac{1 - p \, (1-p) \, \rho \, z + p^3 \, (1-p) \, \rho^2 \, z^2}{D_{\rm BC}(z)} ;
\label{G BC p rho BIS}
\end{equation}
$N_{\rm BC}(z)$ will be the numerator of $G_{\rm BC}(z)$ in eq.~(\ref{G BC p rho BIS}).

\begin{figure}[htb]
\hskip1.5cm
\includegraphics[width=0.75\linewidth]{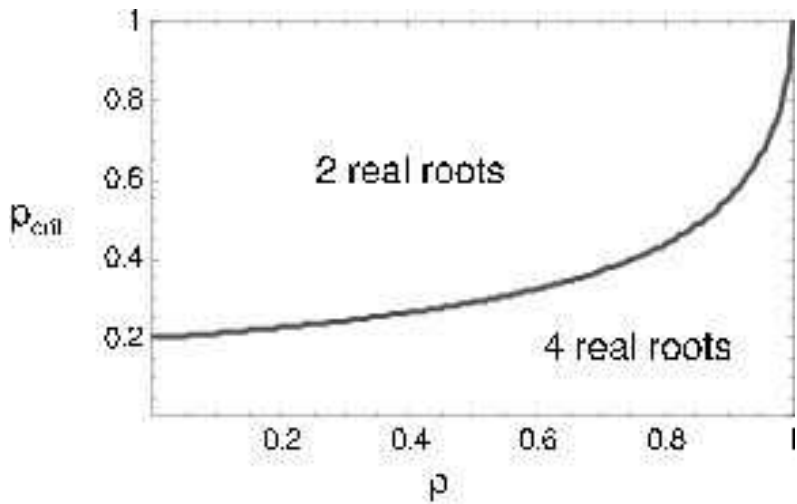}
\vskip0.3cm
\caption{Separation of the domains $(\rho,p)$ for which four real roots exist (under the curve) or two real and two complex roots coexist (above it).}
\label{CourbePCritiqueBC}
\end{figure}

What is the nature of the roots ? Numerically, two situations occur when $0 \leq p \leq 1$ and $0 \leq \rho \leq 1$: there are either (i) four real roots or (ii) two real roots and two complex roots. For instance, when $\rho = 0.9$, there are four real roots if $p < 0.5533938...$, but only two otherwise. The separation of the two domains occurs when one root is degenerate. When this happens, the equalities $P_{\rm carac}^{\rm (BC)}(p,\rho;x) = 0$ and $\displaystyle \frac{\partial P_{\rm carac}^{\rm (BC)}(p,\rho;x)}{\partial x} = 0$ lead to a polynomial constraint satisfied by $p$ and $\rho$, given in the appendix (eq.~(\ref{polynome limite 1})). For a given $\rho$, there exists a unique solution in the range $0 < p < 1$, $p_{\rm crit}(\rho)$, which is displayed in Fig.~\ref{CourbePCritiqueBC}.

\vskip0.5cm

From the inspection of eq.~(\ref{G BC p rho BIS}), we may wonder when numerator and denominator of the generating function may share a common root. A Gr\"{o}bner basis calculation shows that this happens only when $p = 2$, outside our range of interest. Therefore, there are four distinct roots most of the times, and when there is a double (real) root of ${\mathcal P}(p_{\rm crit}(\rho),\rho;x)$, it is indeed degenerate (there is no simplification of $G_{\rm BC}(z)$). The exact expression for ${\rm Rel}_2(S_0 \rightarrow S_n)$ can then be deduced using a partial fraction decomposition of $G_{\rm BC}(z)$. When all the eigenvalues $\lambda_i$ are distinct,
\begin{equation}
G_{\rm BC}(z) = \sum_{\lambda_i} \; \underbrace{\frac{- \lambda_i \, N_{\rm BC}(\frac{1}{\lambda_i})}{D'_{\rm BC}(\frac{1}{\lambda_i})}}_{\textstyle \alpha_i} \; \frac{1}{1 - \lambda_i \, z} ;
\label{decomposition G}
\end{equation}
consequently, the two-terminal reliability reads
\begin{equation}
{\rm Rel}_2(S_0 \rightarrow S_n) = \sum_{\lambda_i} \; \alpha_i \; \lambda_i^n .
\label{decomposition Rel2 formelle}
\end{equation}



In the Brecht-Colbourn case, the root $\lambda_{\rm max}$ of greatest modulus is always real. Unless $p$ and $\rho$ are equal to 1, $\lambda_{\rm max} < 1$, so that ${\rm Rel}_2^{({\rm BC})}(S_0 \rightarrow S_n)$ decreases as $n \rightarrow \infty$, essentially as a power-law behavior
\begin{equation}
{\rm Rel}_2^{({\rm BC})}(S_0 \rightarrow S_n) \sim \lambda_{\rm max}^{n} .
\label{Rel2 asymptotique}
\end{equation}

\subsubsection{Imperfect edges and perfect nodes}


We devote this section to the case of imperfect edges and perfect nodes, since the Brecht-Colbourn ladder was initially investigated in this configuration \cite{BrechtThesis85,BrechtColbourn86}. We immediately find
\begin{equation}
{\rm Rel}_2^{({\rm BC};\rho=1)}(S_0 \rightarrow S_n) = (0 \; 1 \; 0) \; \cdot \; \left(
\begin{array}{c|c|c}
p^2 & 1 & p^2 \\ \hline
p & p & p^2 \\ \hline
- p^2 & - p & p \, (1 - 2 \, p)
\end{array}
\right)^{n} \; \cdot \;
 \left(
\begin{array}{c} 0 \\ 1 \\ 0 \end{array}
\right) ,
\label{Rel2BCparfaits}
\end{equation}
so that the 25-node case \cite{BrechtColbourn86} leads to
\small
\begin{eqnarray}
{\rm Rel}_2(S_0 \rightarrow S_{24}) & = & p^{12}\,\big( 1 + 78\,p + 1121\,p^2 + 6633\,p^3 + 11554\,p^4 - 57525\,p^5 \nonumber \\
& & \hskip1cm - 279450\,p^6 + 89578\,p^7 + 2570040\,p^8 + 1431183\,p^9 \nonumber \\
& & \hskip1cm - 17159566\,p^{10} - 12166498\,p^{11} + 98985590\,p^{12} \nonumber \\
& & \hskip1cm + 33119917\,p^{13} - 495566666\,p^{14} + 212008622\,p^{15} \nonumber \\
& & \hskip1cm + 1867178285\,p^{16} - 2888906214\,p^{17} - 3066846055\,p^{18} \nonumber \\
& & \hskip1cm + 14427083319\,p^{19} - 14178781875\,p^{20} - 17955754991\,p^{21} \nonumber \\
& & \hskip1cm + 80808979717\,p^{22} - 144754404751\,p^{23} + 174732303288\,p^{24} \nonumber \\
& & \hskip1cm - 158925117297\,p^{25} + 113702258108\,p^{26} - 65190712312\,p^{27} \nonumber \\
& & \hskip1cm + 30133254848\,p^{28} - 11197537798\,p^{29} + 3308571601\,p^{30} \nonumber \\
& & \hskip1cm - 761406139\,p^{31} + 131805146\,p^{32} - 16170009\,p^{33} \nonumber \\
& & \hskip1cm + 1254886\,p^{34} - 46368\,p^{35} \big) .
\end{eqnarray}
\label{Rel2 25 noeuds} \normalsize The coefficients are already
quite large, even though the number of nodes remains limited. To the
best of our knowledge, the exact two-terminal reliability polynomial
was calculated only for a 10-node ladder \cite{BrechtThesis85}; we
recover this result by taking $n = 9$ in eq.~(\ref{Rel2BCparfaits}).

\begin{figure}[htb]
\hskip2cm
\includegraphics[width=0.75\linewidth]{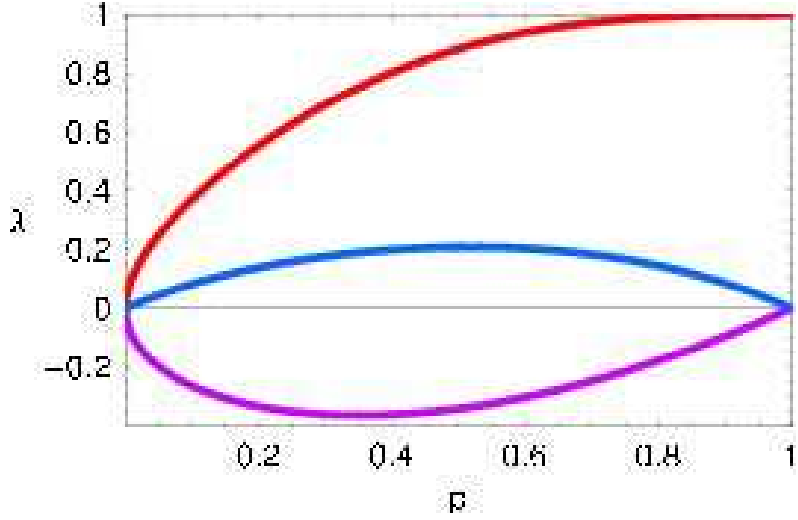}
\vskip0.3cm \caption{Variation with $p$ of the three eigenvalues for the
Brecht-Colbourn ladder with perfect nodes (see eq.~(\ref{ExpressionAnalytiqueLambdas})).}
\label{troisracinesBC}
\end{figure}

The three distinct real eigenvalues $\lambda_i$ ($i = 1,2,3$) of the transfer matrix appearing in eq.~(\ref{Rel2BCparfaits}), one of which is negative, are displayed as a function of $p$ in Fig.~\ref{troisracinesBC}.

\vskip0.0cm

Their analytical expression is ($\xi \in \{1, e^{2 \, i \, \pi/3}, e^{- 2 \, i \, \pi/3}\}$)
\begin{equation}
\lambda = \frac{1}{3} \, p \, (2-p) + \frac{1}{3} \; \left[ \xi \, \left( \frac{{\mathcal A} + i \, \sqrt{27} \, \sqrt{{\mathcal B}}}{2}\right)^{1/3} +
\xi^* \, \left( \frac{{\mathcal A} - i \, \sqrt{27} \, \sqrt{{\mathcal B}}}{2}\right)^{1/3} \right] ,
\label{ExpressionAnalytiqueLambdas}
\end{equation}
where
\begin{eqnarray}
{\mathcal A} & = & - p^2\,\left( 9 - 43\,p + 60\,p^2 - 39\,p^3 + 11\,p^4 \right) , \label{A pour BC} \\
{\mathcal B} & = & {\left( 1 - p \right) }^2\,p^3\,\left( 4 + 9\,p + 16\,p^2 - 88\,p^3 + 98\,p^4 - 32\,p^5 -
8\,p^6 + 5\,p^7 \right) ; \label{B pour BC}
\end{eqnarray}
note that ${\mathcal B}$ is always positive for $0 < p < 1$.

The generating function for the Brecht-Colbourn ladder with perfect nodes is simply
\begin{equation}
G_{\rm BC}^{(\rho=1)}(z) = \frac{1 - p \, (1-p) \, z + p^3 \, (1-p) \, z^2}{1 - p \, (2 - p) \, z - p \, (1-p)^2 \, (1+p) \, z^2 + p^2 \, (1-p)^2 \, z^3} .
\label{G BC p}
\end{equation}
From its partial fraction decomposition, which is necessarily of the form $G_{\rm BC}^{(\rho=1)}(z) = \displaystyle \sum_{i=1}^3 \, \frac{\alpha_i}{1 - \lambda_i \, z}$ because the eigenvalues are distinct, we deduce after simplification
\begin{equation}
{\rm Rel}_2(S_0 \rightarrow S_{n}) = \sum_{i=1}^3 \, \underbrace{\frac{p \, (1-p)^2 - (1-p) \, \lambda_i - \lambda_i^2}{3 \, p \, (1-p)^2 - 2 (1-p)^2 \, (1+p) \, \lambda_i - (2 - p) \, \lambda_i ^2}}_{\textstyle \alpha_i} \; \lambda_i^n .
\label{Rel2 BC rho=1}
\end{equation}
From eq.~(\ref{Rel2 BC rho=1}), it is clear that the large $n$ limit of ${\rm Rel}_2(S_0 \rightarrow S_{n})$ is a power-law behavior, in which the eigenvalue of maximum modulus quickly prevails over the others when $p$ is close to unity, even for moderate values of $n$.

\begin{table}
\caption{Values of the $F_i$'s for the two-terminal reliability polynomial of the 25-node Brecht-Colbourn ladder.}
\vskip2mm
\begin{tabular}{|r|r|r|r|r|r|r|r|} \hline
$i$ & $F_i$ & $i$ & $F_i$ & $i$ & $F_i$ & $i$ & $F_i$ \\ \hline
0    &    1  &  9    &    1125395882 &  18    &    762855455898 &  27    &    3042073238 \\
1    &    47  & 10    &    3974128827  & 19    &    800820887863  &  28    &    635100751 \\
2    &    1079  & 11    &    12199394435  & 20    &    725278875430  &  29    &    105465538  \\
3    &    16103 &  12    &    32708854487 & 21    &    562806091836  &  30    &    13648753 \\
4    &    175418 & 13    &    76833130394 & 22    &    371300292894 &  31    &    1334810   \\
5    &    1484837  &  14    &    158368734141 & 23    &    206539411448  & 32    &    93929  \\
6    &    10151340  & 15    &    286502593795 &  24    &    96061397122  & 33    &    4368  \\
7    &    57524387  & 16    &    454444238576  &  25    &    37052347922  & 34    &    113 \\
8    &    275139029 & 17    &    630595957484  &  26    &    11756780232  & 35    &    1   \\
\hline
\end{tabular}
\vskip1.5cm
\label{ValeursFi25noeuds}
\end{table}

We can also write $\displaystyle {\rm Rel}_2(S_0 \rightarrow S_{n}) = \sum_{i=0}^{2 \, n -1} \, F_i \, p^{2 \, n - 1 - i} \, (1-p)^i$, a well-known expansion of reliability polynomials \cite{Colbourn87}. The $F_i$'s for the 25-node ladder are given in Table~\ref{ValeursFi25noeuds} (the maximum value was equal to 8078 in the 10-node ladder \cite{BrechtThesis85}).
Let us call $m = n+1$ the number of nodes. Using linear regressions on the first values of ${\rm Rel}_2(S_0 \rightarrow S_{n})$, it is straightforward to find
\begin{eqnarray}
{\rm Rel}_2^{(m {\rm \ odd)}}(p) & = & p^{\frac{m-1}{2}} \, (1-p)^{\frac{3 \, m-5}{2}} + \frac{m^2 - 12 \, m -21}{8} \, p^{\frac{m+1}{2}} \, (1-p)^{\frac{3 \, m-7}{2}} \nonumber \\
& & \hskip-2cm + \frac{m^4 +72 \, m^3 +350 \, m^2 - 2376 \, m +2337}{384} \, p^{\frac{m+3}{2}} \, (1-p)^{\frac{3 \, m-9}{2}}
\nonumber \\
& & \hskip-2cm
+ \cdots
\label{autre DL pour m impair}
\end{eqnarray}
\begin{eqnarray}
{\rm Rel}_2^{(m {\rm \ even)}}(p) & = &  \frac{m}{2} \, p^{\frac{m}{2}} \, (1-p)^{\frac{3 \, m-6}{2}} + \frac{(m-2) \, (m^2 +38 \, m +24)}{48} \, p^{\frac{m+2}{2}} \, (1-p)^{\frac{3 \, m-8}{2}} \nonumber \\
& & \hskip-2cm + \frac{(m-2) \, (m^4 +122 \, m^3 +2304 \, m^2 - 5472 \, m -13440)}{3840} \, p^{\frac{m+4}{2}} \, (1-p)^{\frac{3 \, m-10}{2}} \nonumber \\
& & \hskip-2cm + \cdots
\label{autre DL pour m pair}
\end{eqnarray}


Finally, the comparison of the exact results with the various lower bounds proposed in \cite{BrechtColbourn86} is given in Table~\ref{ComparatifBornes} and Fig.~\ref{Comparaison bornes BC 25 noeuds BIS}. It shows that the lower bound of Brecht and Colbourn is rather good for $p$ close to unity, but its sharpness decreases for $p < 0.8$.

\begin{table}
\caption{Comparison of various lower bounds with the exact two-terminal reliability of a 25-node Brecht-Colbourn ladder \cite{BrechtColbourn86}.}
\vskip2mm
\begin{tabular}{|c|c|c|c|c|} \hline
$p$ &   Kruskal-Katona  &   MinCost(edp)    &   Brecht-Colbourn &   exact   \\ \hline
0.75    &   0.031682    &   0.054681    &   0.054681    &   0.625163    \\
0.80    &   0.068803    &   0.119917    &   0.430912    &   0.773696    \\
0.82    &   0.092654    &   0.161200    &   0.558991    &   0.824038    \\
0.84    &   0.124041    &   0.214282    &   0.669269    &   0.867950    \\
0.86    &   0.165305    &   0.281396    &   0.761945    &   0.905042    \\
0.88    &   0.219694    &   0.364529    &   0.837486    &   0.935251    \\
0.90    &   0.291856    &   0.464826    &   0.896659    &   0.958806    \\
0.91    &   0.336579    &   0.521297    &   0.920440    &   0.968231    \\
0.92    &   0.388392    &   0.581555    &   0.940574    &   0.976194    \\
0.93    &   0.448415    &   0.644934    &   0.957259    &   0.982785    \\
0.94    &   0.517724    &   0.710375    &   0.970720    &   0.988109    \\
0.95    &   0.597041    &   0.776313    &   0.981207    &   0.992275    \\
0.96    &   0.686113    &   0.840514    &   0.989003    &   0.995400    \\
0.97    &   0.782518    &   0.899895    &   0.994420    &   0.997608    \\
0.98    &   0.879474    &   0.950274    &   0.997801    &   0.999024    \\
0.99    &   0.961964    &   0.986085    &   0.999524    &   0.999778
\\ \hline
\end{tabular}
\label{ComparatifBornes}
\end{table}

\setlength{\unitlength}{0.4mm}
\linethickness{0.8mm}

\vskip0.0cm
\begin{figure}[htb]
\hskip2.0cm
\includegraphics[width=0.75\linewidth]{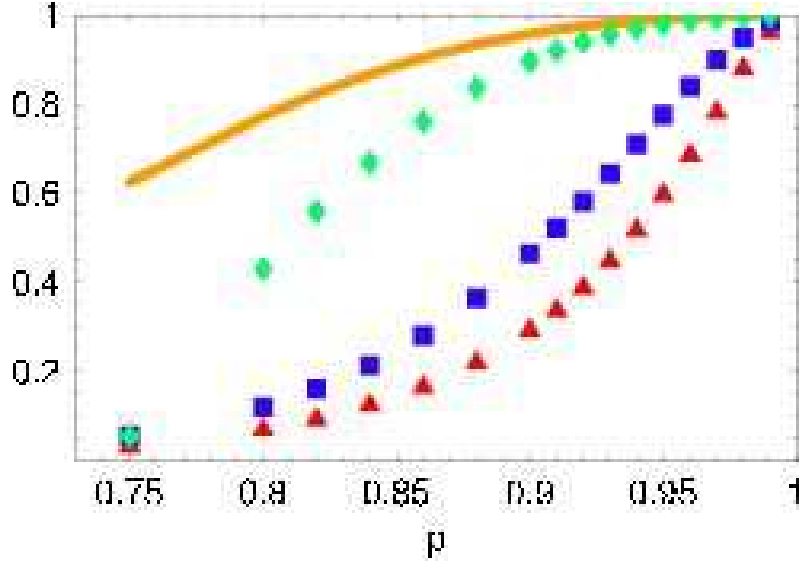}
\vskip0.5cm
\caption{Comparison of the exact value (\protect\textcolor{orange}{\protect\begin{picture}(20,2) \protect\put(0,1){\protect\line(1,0){20}} \protect\end{picture}}) with lower bounds given in ref.~\cite{BrechtColbourn86} for the 25-node ladder:(\textcolor{red}{$\blacktriangle$}) Kruskal-Katona (\textcolor{blue}{$\blacksquare$}) Min Cost (\textcolor{green}{$\blacklozenge$}) Brecht-Colbourn.}
\label{Comparaison bornes BC 25 noeuds BIS}
\end{figure}
\vskip0.5cm


\subsubsection{Imperfect nodes and perfect edges}

The configuration of imperfect nodes and perfect edges, although less studied than the previous one, has nonetheless been considered in several papers \cite{AboElFotoh89,Boesch91,Graver05}. The Brecht-Colbourn ladder case has been studied in detail by Graver and Sobel \cite{Graver05}. The relevant transfer matrix is now equal to
\begin{equation}
M(p=1,\rho) =
\left(
\begin{array}{l|l|l|l}
0 & 0 & \rho & 0 \\ \hline
\rho & \rho & 0 & \rho \\ \hline
1 & \rho & \rho & \rho\\ \hline
- \rho & - \rho & - \rho & - \rho
\end{array}
\right) .
\label{M(p,rho) BC noeuds imparfaits liens parfaits}
\end{equation}
Its characteristic polynomial is $x^2 \, (x^2 - \rho \, x - \rho \, (1 - \rho))$, which simply leads to the eigenvalues 0 --- which plays no role in the final expression of the two-terminal reliability --- and $\displaystyle \lambda_{\pm} = \frac{1}{2} \, \left(\rho \pm \sqrt{4 \, \rho - 3 \, \rho^2} \right)$. The generating function is found to be
\begin{equation}
G_{{\rm BC}}^{(p=1)}(z) = \frac{\rho}{1 - \rho \, z - \rho \, (1 - \rho) \, z^2} = \frac{\rho}{\lambda_+ - \lambda_-} \, \left(\frac{\lambda_+}{1 - \lambda_+ \, z} - \frac{\lambda_-}{1 - \lambda_- \, z} \right) ,
\end{equation}
which gives
\begin{equation}
{\rm Rel}_2(S_0 \rightarrow S_n) = \frac{\rho}{\sqrt{4 \, \rho - 3 \, \rho^2}} \, \left(\lambda_+^{n+1} - \lambda_-^{n+1} \right) .
\end{equation}
This is almost Graver and Sobel's result \cite{Graver05}, obtained through a combinatorial argument. Their final expression differs from ours because their source and destination (in our notation, $S_0$ and $S_n$) are perfect.

\subsection{Generalized fan}

When the reliabilities are $p$ et $\rho$, the transfer matrix of eq.~(\ref{MatriceTransfertEventail}) is equal to
\begin{equation}
\widehat{M}(p,\rho) =
\left(
\begin{array}{l|l|l|l}
p \, \rho  & 0 & 0 & 0 \\ \hline
0 & p \, \rho & p \, \rho & p^2 \, \rho \\ \hline
p^2 \, \rho & p^2 \, \rho & 1 & p^2 \, \rho \\ \hline
-p^2 \, \rho & - p^2 \, \rho & - p \, \rho & p \, (1 - 2 \, p) \, \rho
\end{array}
\right)
\label{M(p,rho) Fan noeuds imparfaits}
\end{equation}
and
\begin{equation}
{\rm Rel}_2(S_0 \rightarrow S_n) = \frac{1}{p} \, (1 \; \rho \; 0 \; 0) \, \cdot \, \widehat{M}(p,\rho)^{n+1} \, \cdot \,
\left(
\begin{array}{c}
1 \\
0 \\
0 \\
0
\end{array}
\right) ,
\label{Rel2finalFanPRho}
\end{equation}
the $1/p$ prefactor being a consequence of the condition $a_0 \equiv 1$ in $\widehat{M}_0$.
The characteristic polynomial of this matrix factorizes nicely:
\begin{equation}
P_{{\rm carac}}^{\rm(fan)}(p,\rho;x) = (x-1) \, (x - p \, \rho) \, \bigg(x - p \, (1-p) \, \rho \bigg)^2 ,
\label{P caracteristique Fan noeuds imparfaits}
\end{equation}
so that the eigenvalues are 1, $p \, \rho$ and $p \, (1-p) \, \rho$, the latter being of degree 2. The presence of 1 among the roots should not be surprising. Indeed, even when $n$ goes to infinity, the two-terminal reliability between $S_0$ and $S_n$ is larger than $p \, \rho^2$, in stark contrast to the Brecht-Colbourn case, where the reliability vanishes. Since a power-law behavior is still expected, the only possibility is that the largest eigenvalue is of modulus 1.

The generating function $G^{({\rm fan})}(z)$ is derived using the recipe described in the Brecht-Colbourn case. The additional convention ${\rm Rel}_2(S_0 \rightarrow S_0) = \rho$ leads to
\begin{eqnarray}
G^{({\rm fan})}(z) & = & \frac{N^{({\rm fan})}(z)}{D^{({\rm fan})}(z)} ,
\label{G fan formel} \\
N^{({\rm fan})}(z) & = & 1 - z \, \big[ 1 + p \, \rho \, (1-p) \, (2 - p \, \rho) \big] + p \, \rho \, z^2 \, \big[ (2 + p \rho)  \, (1-p) \nonumber \\
& & + p^2 \, \rho \, ( p - \rho) \big] - p^2 \, (1-p)^2 \, \rho^2 \, z^3 , \label{Numerateur G Fan}\\
D^{({\rm fan})}(z) & = & (1-z) \, (1 - p \, \rho \, z) \, \big(1 - p \, (1-p) \, \rho \, z\big)^2 . \label{Denominateur G Fan}
\end{eqnarray}
Because all the roots have simple expressions, we expect to find a simple, analytical expression for ${\rm Rel}_2^{({\rm fan})}(S_0 \rightarrow S_n)$. The partial fraction decomposition of eq.~(\ref{G fan formel}) gives indeed
\begin{eqnarray}
G^{({\rm fan})}(z) & = & \frac{\rho^2 \, (1 - p)}{\big(1 - p \, (1-p) \, \rho\big)^2} \, \frac{1- 2 \, p \, (1-p) \, \rho + p^2 \, \rho^2 \, (1 - 3 \, p + p^2)}{1 - p \, (1-p) \, \rho \, z}\nonumber \\[2mm]
& & + p \, \rho^2 \, \frac{1 - p \, \rho \, (2-p)}{1 - p \, (1-p) \, \rho} \; \frac{1}{(1 - p \, (1-p) \, \rho \, z)^2} \nonumber \\[2mm]
& & + \frac{p^2 \, \rho^3}{\big( 1 - p \, (1-p) \, \rho \big)^2} \; \frac{1}{1-z} + \rho \, (1-\rho) \, \frac{1}{1 - p \, \rho \, z} .
\label{GFanElementsSimples}
\end{eqnarray}
The new feature of eq.~(\ref{GFanElementsSimples}) is the $(1 - p \, (1-p) \, \rho \, z)^{-2}$ term. Because
\begin{equation}
\frac{1}{(1- \lambda \, z)^2} =  \sum_{n=0}^{\infty} \, (n+1) \, \lambda^n \, z^n ,
\label{G pole degre 2}
\end{equation}
the final expression of the two-terminal reliability is
\begin{eqnarray}
{\rm Rel}_2^{\rm (fan)}(S_0 \rightarrow S_n) & = & + p^n \, (1-p)^n \, \rho^{n+2} \, \left( n \, p \, \frac{ 1 - p \, \rho \, (2-p)}{1 - p \, (1-p) \, \rho} \right. \nonumber \\
& & \hskip2cm \left. + \frac{1 - p \, \rho \, (2-p) + p^2 \, (1-p)^2 \, \rho^2}{\big(1 - p \, (1-p) \, \rho\big)^2} \right) \nonumber \\
& & + p^n \, \rho^{n+1} \, (1- \rho) + \frac{p^2 \, \rho^3}{\big(1 - p \, (1-p) \, \rho\big)^2} ,
\label{Rel2 Fan}
\end{eqnarray}
where $n$ appears in a prefactor, not only as an exponent. The second term of eq.~(\ref{Rel2 Fan}) is the asymptotic limit when $n \rightarrow \infty$. We see that the reliability of the path $S_0 \rightarrow T \rightarrow S_n$ is enhanced by $1/\big(1 - p \,(1-p) \, \rho\big)^2$. When nodes are perfect, $\rho$ must be set to one in eq.~(\ref{Rel2 Fan}), and the contribution of the eigenvalue $p \, \rho$ vanishes (the transfer matrix of eq.~(\ref{MatriceTransferEventailNoeudsParfaits}) is $3 \times 3$). This leads to
\begin{eqnarray}
{\rm Rel}_2^{({\rm fan};\rho=1)}(S_0 \rightarrow S_n) & = & p^n \, (1-p)^{n+2} \, \left( \frac{ n \, p}{1 - p \, (1-p)} + \frac{(1+ p^2)}{\big(1 - p \, (1-p)\big)^2} \right) \nonumber\\
& & + \frac{p^2}{\big(1 - p \, (1-p) \big)^2} .
\label{Rel2Fan noeuds parfaits}
\end{eqnarray}

\section{Zeros of the two-terminal reliability polynomials}
\label{Zeros}

The structure of the different reliability polynomials may be understood by studying the locations of their zeros in the complex plane. Such a study has been fruitfully performed in the case of the chromatic polynomial \cite{Biggs76,Biggs01,Salas01}, most notably in the context of the four-color theorem. In reliability studies, some effort has been done to discover general properties for the all-terminal reliability ${\rm Rel}_A(p)$ \cite{Chari97,Colbourn93,Oxley02}, its main byproduct being the Brown-Colbourn conjecture \cite{BrownColbourn92}, according to which all the zeros are to be found in the region $|1-p| < 1$. Although valid for series-parallel graphs, this remarkable conjecture does not strictly hold in the general case (but not by far) \cite{Royle04}. As mentioned in the introduction, the all-reliability polynomial is linked to the Tutte polynomial, an invariant of the graph. It has also been studied extensively by Chang and Shrock for various recursive families of graphs \cite{Chang03}, who give the limiting curves where all zeros of the polynomials converge.


In this section, after briefly recalling general results of the literature, we study the roots of ${\rm Rel}_2(p,\rho) = 0$ in the complex plane for a fixed $\rho$, in order to see whether some insight may also be gained in this case. Admittedly, this polynomial depends on the couple (source, terminal), but structures are still expected. We show that the zeros tend to aggregate along portions of algebraic curves, which can substantially differ even for the two families of graphs under consideration, even though they have the same all-terminal reliability. Moreover, structural transitions occur at $\rho = \frac{1}{2}$.

\subsection{Calculation of the limiting curves (generalities)}

As $n$ grows, the number of zeros of the reliability polynomial in the complex plane increases. Because of the matrix transfer property, we have recursion relations between relability polynomials corresponding to successive values of $n$. The general treatment of the problem has been done by Beraha, Kahane, and Weiss \cite{Beraha78}, but may be understood in the following, simplifying way: if the reliability polynomial is of the form $\sum_i \alpha_i \, \lambda_i(p)^n$ (where $\lambda_i$ are the eigenvalues of the transfer matrix), then at large $n$, only the two eigenvalues of greater modules, say $\lambda_1$ and $\lambda_2$, will prevail, so that the reliability polynomial will vanish when $|\lambda_1(p)| = |\lambda_2(p)|$ (of course, it might be three or more eigenvalues of equal modulus; the present oversimplification works quite well here). This defines a set of curves in the complex plane, where all zeros should accumulate in the $n \rightarrow \infty$ limit. The interested reader should refer to the work of Salas and Sokal \cite{Salas01} for a very detailed discussion of the convergence to the limiting curves. This behavior is not modified when one or more of the $\alpha_i$'s is a polynomial in $n$ \cite{Beraha78}, as will become apparent for the generalized fan.


\subsection{Calculation for the Brecht-Colbourn ladder}

The recursion relation obeyed by ${\rm Rel}_2^{({\rm BC})}(S_0 \rightarrow S_n) \equiv {\rm Rel}_2^{(n)}$ is easily deduced from eq.~(\ref{P caracteristique BC noeuds imparfaits}); it reads
\begin{eqnarray}
{\rm Rel}_2^{(n)}(p) & = & p \, (2 - p) \, \rho \, {\rm Rel}_2^{(n-1)} + p \, \rho \, (1 - p \, \rho - p^2 \, \rho + p^3 \, \rho) \, {\rm Rel}_2^{(n-2)} \nonumber \\
& & \hskip-1cm - p^2 \, (1-p) \, \rho^2 \, (1-\rho) \, {\rm Rel}_2^{(n-3)} + p^4 \, (1-p) \, \rho^3 \, (1-\rho) \, {\rm Rel}_2^{(n-4)}.
\end{eqnarray}
and leads to very quick calculations using mathematical softwares such as Mathematica \cite{Mathematica}. Note that this recursion relation of order four becomes of order three when $\rho =1$. We can then look for solutions of ${\rm Rel}_2^{(n)}(p) = 0$ in the complex plane. Of course, the degree of the polynomial, as well as the magnitude of its coefficients, will increase with $n$.

\subsubsection{Perfect nodes}

\begin{figure}
\hskip1.5cm
\includegraphics[width=0.75\linewidth]{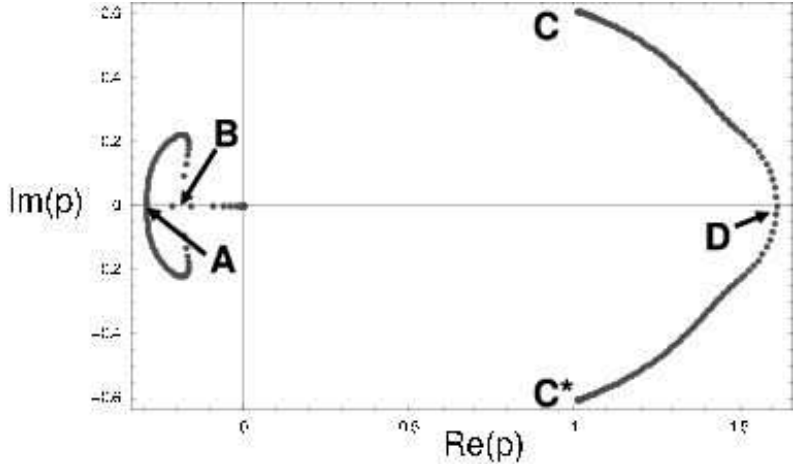}
\caption{Localization of the complex roots $p$ of the two-terminal reliability polynomial for a
Brecht-Colbourn ladder of 151 perfect nodes and 299 edges.}
\label{BC151}
\end{figure}

Let us begin by setting $\rho =1$. A first step is to display the roots of ${\rm Rel}_2^{(n)}(p) = 0$. These zeros have been calculated using the \verb?NSolve[]? routine of Mathematica; because the polynomial coefficients can be very large, a numerical accuracy of several hundred digits is sometimes necessary. Figure~\ref{BC151} shows the location of the zeros of ${\rm Rel}_2^{(150)}(p)$ in the complex plane (it corresponds to a graph with 151 nodes and 299 edges).


On the right half of the complex plane, we have a simple, open curve crossing the real axis at $D$ and extremities at $C$ and $C^{*}$. On the left half, however, we have what looks like a closed curve --- intersecting the negative real axis in $A$ and $B$ --- as well as some dots on the negative real axis, between $A$ and the origin.

As mentioned above, it is well known that as $n$ increases, the zeros accumulate at particular locations, constituted by segments/portions of algebraic curves. For $n = 150$, these limits are almost reached, even though the ``sampling'' is not uniform. We can calculate the limiting curves by considering that the roots of the characteristic polynomial are $\lambda \, e^{i \, \theta/2}$, $\lambda \, e^{- i \, \theta/2}$ and $\mu$. The coefficients of the characteristic polynomial imply constraints between $\lambda$ (which may be complex), $\mu$ and $\theta$, or more precisely, $t = \cos \frac{\theta}{2}$ or $T = \cos \theta$, which will be used in the following. For instance, $\mu = (2-p) \, p - 2 \, \lambda \, \cos \frac{\theta}{2}$. After eliminating $\lambda$ and $\mu$, we obtain a compatibility condition which must be satisfied by $p$, and $T$ (the polynomial ${\mathcal P}_3(p,\rho,T)$ is given in eq.~(\ref{polynome critique P3})):



\begin{eqnarray}
{\mathcal P}_3(p,\rho=1,T) & = & 2 + 9 \, p -4 \, p^2 -30 \, p^3 +38 \, p^4 -10 \, p^5 -6 p^6 +3 \, p^7 \nonumber \\
& & +\left(2 +12 \, p -8 \, p^2 -34 \, p^3+48 \, p^4 -18 \, p^5 -2 \, p^6 +2 \, p^7\right) \, T \nonumber \\
& &  -4 \, p \, (1-p)^2 \, \left(1 - p +p^2\right) \, T^2 -8 \, p \, (1-p)^2 \, T^3 \nonumber \\
& = & 0.
\label{P3rhoegal1}
\end{eqnarray}
The additional requirement $|\mu| < |\lambda|$
translates into an additional contraint between $p$ and $T$.

A plot of this parametric set of curves shows indeed that the whole portion of the negative real axis between $A$ and the origin is indeed a solution, so that we could expect more zeros there when $n > 150$.

A few critical points can also be deduced from eq.~(\ref{P3rhoegal1}). For instance, $D$ corresponds to a real solution of ${\mathcal P}_3(p,\rho=1,T=-1) = 0$, namely $p_D = \frac{1+\sqrt{5}}{2}$. $C$ and $C^{*}$ are points of the complex plane defined by $p_C$ and its complex conjugate, which are roots of ${\mathcal P}_3(p,\rho=1,T = +1) = 0$, that is
\begin{equation}
4 + 9 \, p + 16 \, p^2 - 88 \, p^3 + 98 \, p^4 - 32 \, p^5 - 8 \, p^6 + 5 \, p^7 = 0 .
\label{valeurpCrhoegal1}
\end{equation}
This polynomial already appeared in the expression of ${\mathcal B}$. Note that there are more than one pair of complex solutions to this equation, the relevant one is given by $p_C \approx 1.011578 + i \, 0.607394$. The determination of the complex numbers $p_A$ and $p_B$ associated with $A$ and $B$ is a little more elaborate. Both $p_A$ and $p_B$ are real and negative; they correspond to roots of degree two of eq.~(\ref{P3rhoegal1}). In order to find them, we must have ${\mathcal P}_3(p,\rho=1,T) = 0$ and  $\displaystyle \frac{\partial {\mathcal P}_3(p,\rho=1,T)}{\partial p} = 0$. These two relations are satisfied for special values of $p$ (or $T$). After substitution and elimination of $T$ using Mathematica, we obtain a product of polynomials, which must cancel for $p = p_A$ and $p = p_B$. For the sake of completeness, we provide the needed polynomials in appendix \ref{p1 p2 p3}. A close numerical study allows to identify the relevant polynomials:
\begin{itemize}
\item $p_A \approx -0.2879878$ is solution of ${\mathcal P}_1(p, \rho=1) = 0 $ (see eq.~(\ref{polynome critique P1})), that is
\begin{equation}
0 = 1 - p - 11\,p^2 + 15\,p^3 - 3\,p^4 - 2\,p^5 - p^6 + p^7 ,
\label{valeurpArhoegal1}
\end{equation}
attained for $T \approx 0.138176$, a root of
\begin{equation}
0 = -1 + 11\,T - 27\,T^2 - 26\,T^3 + 140\,T^4 + 240\,T^5 + 144\,T^6 + 32\,T^7.
\label{valeurTArhoegal1}
\end{equation}

\item $p_B \approx -0.1849482$ is solution of ${\mathcal P}_2(p, \rho=1) = 0$  (see eq.~(\ref{polynome critique P2})), i.e.,
\begin{eqnarray}
0 & = & 1 - 4\,p - 9\,p^2 + 101\,p^3 - 413\,p^4 + 1019\,p^5 - 1761\,p^6 \nonumber \\
& & + 2151\,p^7 - 1864\,p^8 + 1097\,p^9 - 386\,p^{10} + 60\,p^{11} ,
\label{valeurpBrhoegal1}
\end{eqnarray}
attained for $T \approx -0.9511957$, one of the solutions of
\begin{eqnarray}
0 & = & 3040707 + 12576464\,T + 15322821\,T^2 - 4376828\,T^3 \nonumber \\
& & - 22559186\,T^4 - 10112842\,T^5 + 9510320\,T^6 + 7762048\,T^7 \nonumber \\
& & - 1337920\,T^8 - 2068608\,T^9 + 8192\,T^{10} + 204800\,T^{11}
\label{valeurTBrhoegal1}
\end{eqnarray}
\end{itemize}

\subsubsection{Imperfect nodes}

We can perform the study for $\rho < 1$ by following the same lines as above. Although more cumbersome because there are now four possible roots instead of three, the calculations are interesting nonetheless because we may vary one parameter, $\rho$. A natural question is: can it affect the global structure of the zeros ? The answer is yes, and the location of the zeros undergoes a kind of ``structural transition'' for a particular value of $\rho$, namely $\frac{1}{2}$.

Let us first consider $\rho > \frac{1}{2}$. The location of the zeros is then qualitatively similar to that described in Fig.~\ref{BC151}, the whole structure merely expanding from the origin. After comparison with the numerical values for $n=150$, $D$ appears to be associated with $p_D = \frac{1}{2} \, (1+\sqrt{1+\frac{4}{\rho}})$, the relevant solution of ${\mathcal P}_3(p, \rho, T = -1) = 0$.  Likewise, $C$ and $C^{*}$ are given by $p_C$, which is one complex solution of ${\mathcal P}_3(p, \rho, T = +1) = 0$ (${\mathcal P}_3$ is given in eq.~(\ref{polynome critique P3})).

The determination of $p_A$ and $p_B$ is more tedious, because ${\mathcal P}_3(p, \rho, T)$ is now a polynomial of degree 6 in $T$. Here again, we must find the common zeros of ${\mathcal P}_3(p, \rho, T)$ and its derivative with respect to $p$. The elimination of $T$, performed using Mathematica, leads to a polynomial in $p$ and $\rho$, which must vanish. Actually, this polynomial can be factored, and a numerical comparison with the zeros obtained for $n=150$ shows that $p_A$ is solution of ${\mathcal P}_1(p, \rho) = 0$ (see eq.~(\ref{polynome critique P1})). Similarly, $p_B$ is a solution of ${\mathcal P}_2(p, \rho) = 0$, where ${\mathcal P}_2$ is given in eq.~(\ref{polynome critique P2}).

For $\rho = \frac{1}{2}$, $p_A \approx -0.4359355$, $p_B \approx -0.2885759$, $p_C \approx 0.748541 + i \, 1.03759$, and $p_D = 2$. Note that $p_A$ and $p_B$ are zeros of polynomials in $p$ of degrees equal to 22 and 30, respectively (having $\rho \neq 1$ does not simplify the problem).

\begin{figure}
\hskip1cm
\includegraphics[width=0.75\linewidth]{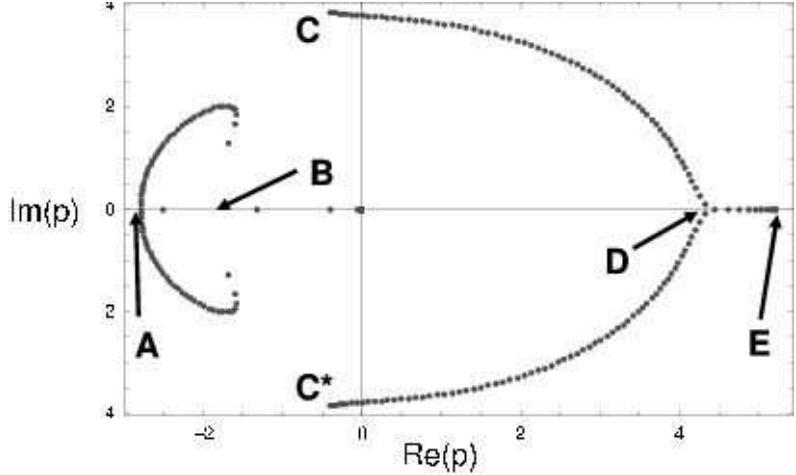}
\caption{Localization of the complex roots $p$ of the two-terminal reliability polynomial for a
Brecht-Colbourn ladder of 151 nodes and 299 edges, when $\rho = 10^{-2}$.}
\label{BC151Rho0.01}
\end{figure}

For $\rho < \frac{1}{2}$, the above expressions for $p_A$, $p_B$, and $p_C$ are still valid, even though $C$ and $C^{*}$ may now belong to the left half-plane. However, on the right half-plane, the structure of zeros undergoes a drastic transformation, as shown in Fig.~\ref{BC151Rho0.01} for $\rho = 10^{-2}$.
$D$ is now an angular point, and zeros may be found on the real axis between $D$ and $E$. After numerical comparisons between structures of zeros and limiting curves have been performed, it appears that $p_D$ is now a real root of ${\mathcal P}_1(p, \rho) = 0$ (actually, the third one, in decreasing order), while $p_E$ is another real root of ${\mathcal P}_3(p, \rho, T = +1) = 0$.

\subsubsection{Asymptotic limits when $\rho \rightarrow 0$}

In this section, we address the asymptotic dependence as $\rho$ decreases to zero of all the critical points $A$, $B$, $C$, $D$, and $E$. A first step consists in finding the corresponding $p$'s for (very) small values of $\rho$. It is not too difficult to observe that these numerical values have a $\rho^{-1/3}$ dependence. We can then use improve on this knowledge by using the algebraic equations of which they are solutions, making again use of Mathematica for the asymptotic expansions. A comparison with the numerical results determines the true leading term of the expansion; after some work, we obtain
\begin{eqnarray}
p_A(\rho) & = & - \left( \frac{3 - \sqrt{5}}{2 \, \rho}\right)^{1/3} + \frac{35 + 12 \, \sqrt{5}}{90} + O(\rho^{1/3}) ,\\
p_B(\rho) & = & - \left( \frac{\chi}{\rho}\right)^{1/3} + \alpha +  O(\rho^{1/3}) ,
\end{eqnarray}
where
\begin{equation}
\chi = \frac{1}{15} \; \left\{ \left( \frac{2531 + 15 \, \sqrt{31593}}{2}\right)^{1/3} - \left(
\frac{-2531 + 15 \, \sqrt{31593}}{2}\right)^{1/3} - 8 \right\}
\end{equation}
is the real root of $5 \, \chi^3 + 8 \, \chi^2 + 8 \, \chi - 1 = 0$. Numerically, $\chi \approx
0.11166155366$ and $\kappa = \chi^{1/3} \approx 0.48154242495$.  The next-order calculation provides
\begin{eqnarray}
p_B(\rho) & = & - \frac{\kappa}{\rho^{1/3}} + \underbrace{\frac{1}{50} \, \frac{70147451 \, \kappa^2+22890531 \, \kappa -3689542}{1685743 \, \kappa ^2+778683 \, \kappa -121256}}_{\textstyle \alpha} + O(\rho^{1/3}) \\[3mm]
& \approx & - \frac{0.48154242495}{\rho^{1/3}} + 0.38969988720 + O(\rho^{1/3}) .
\end{eqnarray}
Proceeding in a similar way for the critical points defining the rightmost structure in the complex plane, we find
\begin{eqnarray}
p_D(\rho) & = & \left( \frac{3 - \sqrt{5}}{2 \, \rho}\right)^{1/3} + \frac{19 + 4 \, \sqrt{5}}{30} + O(\rho^{1/3}) \\
p_E(\rho) & = & \left( \frac{3 - \sqrt{5}}{2 \, \rho}\right)^{1/3} + \frac{2}{15} \, \left( \frac{3 + \sqrt{5}}{2}\right)^{5/6} \, \sqrt{-75 + 35 \, \sqrt{5}} \, \frac{1}{\rho^{1/6}} \nonumber \\
& & + \frac{11 + 4 \, \sqrt{5}}{30} + O(\rho^{1/6}) \\
p_C(\rho) & = & e^{2 \, i \, \pi/3} \, \left( \frac{3 - \sqrt{5}}{2 \, \rho}\right)^{1/3} +  \frac{2}{15} \, \left( \frac{3 + \sqrt{5}}{2}\right)^{5/6} \, \sqrt{-75 + 35 \, \sqrt{5}} \; \frac{e^{i \, \pi/3} \,}{\rho^{1/6}} \nonumber \\
& & + \frac{11 + 4 \, \sqrt{5}}{30} + O(\rho^{1/6})
\end{eqnarray}
Note that the expansion $p_C(\rho)$ is nothing but $p_E(\rho)$, with the transformation $\rho \longrightarrow e^{- 2 \, i \, \pi} \; \rho$. It corresponds to the second complex root, by decreasing order of the real part.

\subsection{Calculation for the fan}

We expect here much easier calculations, since all the eigenvalues are extremely simple: 1, $p \, \rho$, and $p \, (1-p) \, \rho$.  They are, indeed, since the limiting curves can be expressed analytically (see Table~\ref{LimitesEventail}). It is worth noting that nonetheless, other structure transitions occur at $\rho = 1$ and $\rho = \frac{1}{2}$. Using  the relevant recursion relation between successive reliability polynomials, we can again calculate ${\rm Rel}_2^{{\rm (fan)}}(S_0 \rightarrow S_n)(p)$ rapidly using Mathematica.


\vskip0.5cm
\begin{figure}[htb]
\hskip1.0cm
\includegraphics[width=0.75\linewidth]{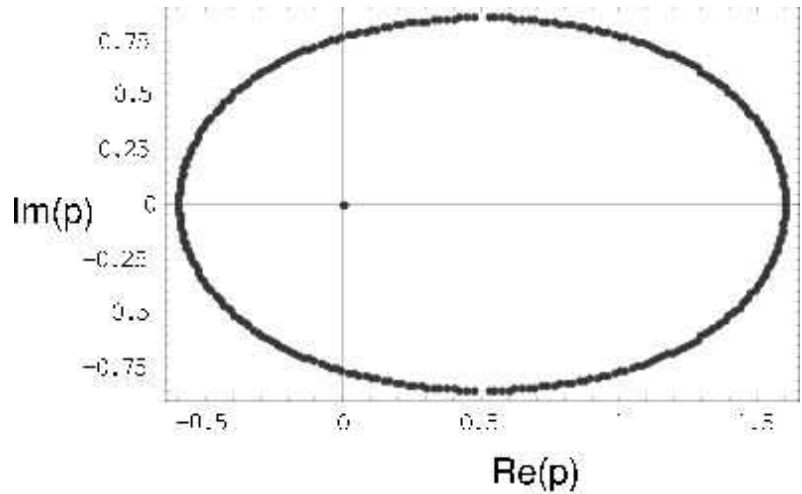}
\vskip0.0cm \caption{Location of the roots of ${\rm Rel}_2^{(n=150)}(p) = 0$ for the fan, with $\rho = 1$.}
\label{Zeros Fan 152 noeuds rho egale 1}
\end{figure}

For $\rho = 1$, we see on Figure~\ref{Zeros Fan 152 noeuds rho egale 1} that the limiting curve is now essentially a closed one, even though the zeros appear to belong to two different sets which slowly join as $n$ increases. The presence of $n$ in a prefactor as in eqs.~(\ref{Rel2 Fan}) and (\ref{Rel2Fan noeuds parfaits}) does not affect the general definition of the limiting curves \cite{Beraha78}. The limiting curve is very easy to obtain, since we have only two eigenvalues, 1 and $p \, (1-p)$. For them to have the same modulus implies $p \, (1-p) = - e^{i \, \theta}$, so that we get $p = \frac{1}{2} \, \left( 1 \pm \sqrt{1 + 4 \, e^{i \, \theta}} \right)$.

\begin{figure}[htb]
\hskip1.0cm
\includegraphics[width=0.75\linewidth]{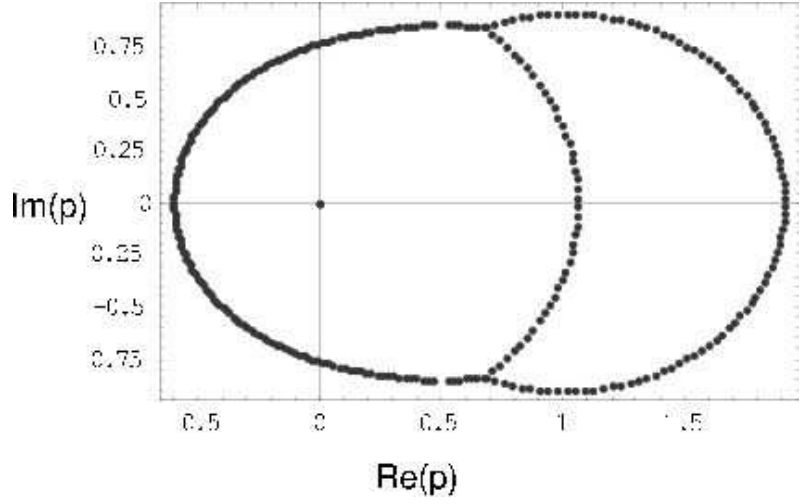}
\vskip0.0cm \caption{Location of the roots of ${\rm Rel}_2^{(n=150)}(p) = 0$ for the fan, with $\rho = 0.9999$.}
\label{Zeros Fan 152 noeuds rho egale 0.9999}
\end{figure}

When $\rho$ is strictly less than 1, the whole structure is altered, as witnessed by Figure~\ref{Zeros Fan 152 noeuds rho egale 0.9999} for $\rho = 0.9999$. Portions of circles centered at the origin and at the point $(1,0)$ are added. The reason for such additional structures comes from the existence of a new eigenvalue $p \, \rho$. They persist down to $\rho = \frac{1}{2}$. For $\rho < \frac{1}{2}$, the only eigenvalues that matter are actually 1 and $p \, (1-p) \, \rho$, so that the structure looks again very much like Fig.~\ref{Zeros Fan 152 noeuds rho egale 1}.

\vskip0.5cm

\begin{table}
\caption{Analytical expressions of the limiting aggregation curves for the location of $p$ such that ${\rm Rel}_2^{{\rm (fan)}, n \rightarrow \infty}(p,\rho) = 0$, as a function of $\rho$, the node reliability.}
\vskip2mm
\begin{tabular}{|c|l|l|} \hline
& & \\[-6mm]
node reliability & parametrization & validity range \\
& & \\[-12mm]
& & \\ \hline
& & \\[-5mm]
$\rho = 1$ & $p = \frac{1}{2} \, \left( 1 \pm \sqrt{1 + 4 \, e^{i \, \theta}} \right)$ & $-1 \leq \cos  \theta \leq 1$ \\[-5mm]
& & \\ \hline
& & \\[-5mm]
$\frac{1}{2} < \rho < 1$  &
$\displaystyle \begin{array}{l}
p = \frac{1}{2} \, \left( 1 - \sqrt{1 + \frac{4}{\rho} \, e^{i \, \theta}} \right) \\[2mm]
p = \frac{1}{2} \, \left( 1 + \sqrt{1 + \frac{4}{\rho} \, e^{i \, \theta}} \right) \\[2mm]
p = \frac{1}{\rho} \, e^{i \, \theta} \\[2mm]
p = 1 + e^{i \, \theta}
\end{array} $
 &
$\displaystyle
\begin{array}{l}
-1 \leq \cos  \theta \leq 1 \\[2mm]
-1 \leq \cos  \theta \leq \frac{1}{2 \, \rho} \, \left( \frac{1}{\rho^2} -3\right) \\[2mm]
\cos \theta \geq \frac{1}{2 \, \rho} \\[2mm]
\cos \theta \geq \frac{1}{2 \, \rho^2} -1
\end{array}
$ \\
& & \\[-12mm]
& & \\ \hline
& & \\[-6mm]
$0 < \rho \leq \frac{1}{2}$ & $p = \frac{1}{2} \, \left( 1 \pm \sqrt{1 + \frac{4}{\rho} \, e^{i \, \theta}} \right)$ & $-1 \leq \cos  \theta \leq 1$ \\[-5mm]
& & \\ \hline
\end{tabular}
\vskip1.5cm
\label{LimitesEventail}
\end{table}

Their derivation being straightforward, we simply give the analytical expressions of the limiting curves in Table~\ref{LimitesEventail}. When $\rho \rightarrow 0$, the structure expands with a scaling factor of $\rho^{-1/2}$, in contrast with the $\rho^{-1/3}$ obtained for the Brecht-Colbourn ladder.

\section{All-terminal reliability}
\label{All-terminal reliability}

As mentioned in the introduction, the all-terminal reliability ${\rm Rel}_A$ is another useful measure of the network availability, giving the probability that all nodes are connected. In this case however, the node reliabilities are a mere overall factor, so that they may be considered equal to 1 for all practical purposes \cite{Colbourn87}. Chang and Shrock \cite{Chang03} gave the explicit expressions of ${\rm Rel}_A$ for various recursive families of graphs, among which the Brecht-Colbourn ladder, with the same reliability $p$ for all edges. The all-terminal reliability of the generalized fan has been calculated by Neufeld and Colbourn \cite{Neufeld85}. The results are of course identical, because at each step, two edges are added, while the common original graph is the triangle (the complete graph $K_3$). For the sake of completeness, we slightly generalize their result for edges with distinct reliabilities. Here again, the
final, analytical expression can be written in a concise form using transfer matrices. In the
context of graph theory, this can be viewed as the factorization of a particular value of the
multi-variate Tutte polynomial, considered by Wu \cite{Wu78} and Sokal \cite{Sokal05}.

\vskip0.5cm
\begin{figure}[htb]
\hskip1.50cm
\includegraphics[width=0.75\linewidth]{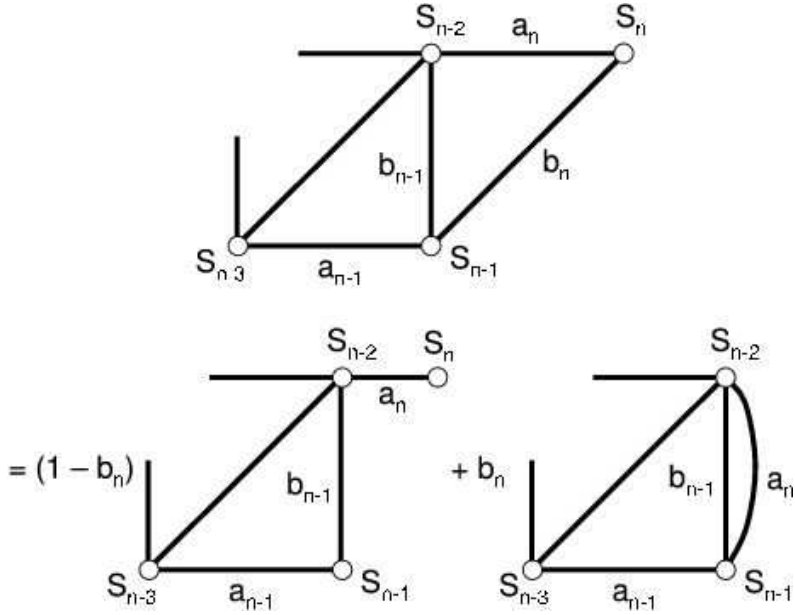}
\vskip0.0cm \caption{Deletion-contraction simplification for the all-terminal reliability of the Brecht-Colbourn ladder.}
\label{SimplificationRelA}
\end{figure}

We expect the all-terminal reliability ${\mathcal R}_n = {\rm Rel}_A(S_0 \leftrightarrow S_n)$ for the Brecht-Colbourn ladder to exhibit the same behavior as the two-terminal reliability, with a transfer matrix depending on $a_n$ and
$b_n$, and a generating function that is a rational fraction of $p$ (the common reliability of
links) and $z$. As mentioned above, this calculation should be somewhat easier because all nodes can
be considered perfect without loss of generality \cite{Colbourn87}. We can then use the usual
pivotal decomposition to establish a relationship between ${\mathcal R}_n$ and ${\mathcal
R}_{n-1}$, as represented in Fig.~\ref{SimplificationRelA}. We find
\begin{equation}
{\mathcal R}_n = (1 - b_n) \, a_n \, {\mathcal R}_{n-1} + b_n \, {\mathcal R}_{n-1} (b_{n-1}
\rightarrow b_{n-1} \, // \, a_n) \label{recurrence pour Rn}
\end{equation}
Using the ansatz ${\mathcal R}_n = \alpha_n \, (a_n + b_n) + \gamma_n \, a_n \, b_n$, we get
\begin{equation}
\left(
\begin{array}{l}
\alpha_n \\
\gamma_n
\end{array}
\right) = \left(
\begin{array}{l|l}
a_{n-1}+ b_{n-1} & a_{n-1} \, b_{n-1} \\ \hline
1 - a_{n-1} - 2 \, b_{n-1} & a_{n-1} \, \big( 1 - 2 \, b_{n-1} \big)
\end{array}
\right) \; \left(
\begin{array}{l}
\alpha_{n-1} \\
\gamma_{n-1}
\end{array}
\right)
\label{matrice de transfert pour Rn}
\end{equation}
with $a_1 = \gamma_1 = 0$ and $\alpha_1 = 1$. Calling $\widetilde{M}_{n-1}$ the transfer matrix of eq.~(\ref{matrice de transfert pour Rn}), we obtain
\begin{equation}
{\mathcal R}_n = (1 , 0) \; \cdot \; \widetilde{M}_n \; \cdot \; \widetilde{M}_{n-1} \; \cdots  \;
\widetilde{M}_1 \;  \cdot \; \left(
\begin{array}{l}
1 \\
0
\end{array}
\right)
\label{expression de Rn}
\end{equation}
When all reliabilities are equal to $p$, the transfer matrix is simply
\begin{equation}
\widetilde{M'} = \left(
\begin{array}{l|l}
2 \, p &  p^2 \\ \hline
1 - 3 \, p & p \, (1 - 2 \, p)
\end{array}
\right)
\label{matrice transfert pour Rn avec p}
\end{equation}

The eigenvalues of $\widetilde{M'}$ are $\displaystyle \zeta_{\pm} = \frac{p}{2} \; \left( 3 - 2 \, p
\pm \sqrt{5 - 8 \, p + 4 \, p^2} \right)$, so that the generating function is
\begin{equation}
G_R(z) = \frac{p \, z^2}{1 - p \, (3 - 2 \, p) \, z + p^2 \, (1-p) \, z^2}
\label{GRn final}
\end{equation}
and\begin{equation}
{\mathcal R}_n = \frac{1}{\sqrt{5 - 8 \, p + 4 \, p^2}} \; \left(\zeta_{+}^{n-1} - \zeta_{-}^{n-1}
\right) \hskip1cm n \geq 2 \label{Rn final}
\end{equation}
The last results were obtained in \cite{Chang03}, while the same expression can be found in the paper by Neufeld and Colbourn \cite{Neufeld85}.

\section{Conclusion and Perspectives}
\label{Perspectives}

We have given the exact solution of the two- and all-terminal reliabilities for the Brecht-Colbourn ladder and the generalized fan for arbitrary size and individual element reliability. While simple, these graphs correspond nonetheless to realistic network architectures, especially in telecommunication networks for IP transport. Node and edge failures are put on an equal footing, and the simple formulae relying on transfer matrices may be directly implemented. We have also given the analytical solution of the two- and all-terminal reliabilities, when reliabilities are $p$ and $\rho$ for edges and nodes, along with their rational generating functions. The locations of the zeros of the two-terminal reliability polynomials differ for the two families of graphs considered in this paper, even though their all-terminal reliability is identical. They possess structures in the ${\rm Re}(p) < 0$ region, and exhibit transitions for particular values of $\rho$.


Even though the delta-star transformation has also been successfully
applied to the case of simple ladders \cite{CTechellesimple}, it may
not be so easy to use in more complicated networks. However, it
seems quite clear that a similar decomposition through transfer
matrices of two-terminal reliabilities should occur for ladders of
greater width too. In order to make such calculations useful for
applications, imperfect nodes as well as imperfect edges must be
considered. All is needed is a recursion relation between successive
graphs, when one ``elementary brick'' is added. This implies a new
expression for the deletion-contraction theorem, in which the
linearity with respect to all individual edge or node reliabilities
must be preserved, whereas --- to our knowledge --- edge
reliabilities are often renormalized to account for the
unreliability of the nodes they connect
\cite{Aggarwal75,Theologou91,Torrieri94}. This new expression will
be given elsewhere, along with an application to other recursive
families of graphs such as the $K_4$ ladder and the $K_3$ cylinder
\cite{CTlettregenerale}.



What should we expect ? Basically, the same kind of behavior as detailed in the present work, with a
factorization of the reliability in terms of transfer matrices, the dimension of which
substantially increase to reflect the interplay of different edges/nodes in the overall
reliability, and the underlying algebraic structure of the graph (one cannot escape the intrinsic complexity of the problem...).


There are obviously many directions in which this work may be further extended, for instance in the estimates of bounds. The first one is to use the present results, or their future extensions \cite{CTlettregenerale,CTpreparation}, as possible upper or lower bounds to more complex graphs. If a graph under consideration looks like --- or made as such --- a special instance of a recursive family of graphs, we expect the generating function to be a rational fraction again. The dimension of the corresponding transfer matrix may be probed by trying to find recursion relations between successive reliability polynomials. Of course, if the dimension of the corresponding transfer matrix is large --- which is most likely to happen --- the degree of the numerator and denominator of the fraction may be too large for a complete solution to be
obtained easily. Even so, the knowledge that the {\em true} generating function is rational may
be an indication that an {\em approximate} generating function might still be quite useful, because
Pad\'{e} approximants\cite{Baker96,NumericalRecipes} are known for their devilish knack of
getting {\em very} close to the exact function. This will be addressed elsewhere \cite{CTpreparation}. Our calculations may also provide some information on some combinatorial issues, such as the enumeration of self-avoiding walks on lattices of restricted width.


Other quantities of interest may be deduced from the general expression of the network reliability, among them the sensitivity of a equipment \cite{CTechellesimple}, the influence of scheduled maintenance on the overall network availability, and the failure frequency \cite{Hayashi91,Singh77}, which can all be deduced from various partial derivatives of the two- and all-terminal reliabilities \cite{CTdispersion}. Since the reliability of each equipment appears in only one transfer matrix, the computation of all these network parameters is straightforward. One should also be aware that the reliability of each equipment is not known with absolute accuracy. The consequence of this uncertainty on the overall reliability has been addressed by Coit and collaborators in the case of series-parallel reducible networks \cite{Coit97,Coit04}. Our transfer matrix factorization makes this topic another instance of a product of random matrices, definitely a vast field in mathematical physics \cite{Crisanti93}.

Finally, the exact results found for classes of arbitrarily large networks may prove useful for
testing different algorithms (Monte Carlo, genetic, OBDD, etc) in numerically exacting
configurations, where edge and node unreliabilities must both be taken into account.

\section*{Acknowledgments}
\label{Acknowledgments}

I am indebted to \'{E}lisabeth Didelet for giving me the opportunity to study reliability issues in
depth, and to Jo\"{e}l Le Meur,
Fran\c{c}ois Gallant, Veluppillai Chandrakumar,
Adam Ouorou, and \'{E}ric Gourdin for encouragement and support. Useful information from Prof. J. Graver is gratefully acknowledged. I would also like to thank Annie Druault-Vicard, Christine Leroy, Herv\'{e} Coutelle, and Gilles Petitdemange for very useful discussions.

\vfill
\eject

\appendix

\section{Polynomials used in the determination of critical points for the zeros of ${\rm Rel}_2(p)$ (Brecht-Colbourn ladder)}
\label{p1 p2 p3}

\subsection{${\mathcal P}_1(p,\rho)$}
\label{plimite1}

\small
\begin{eqnarray}
{\mathcal P}_1(p,\rho) & = & \rho  - p\,\left( 1 + 9\,\rho  +
       3\,{\rho }^2 \right) +
  p^2\,\rho\, \left( 41 + 37\,\rho  - 11\,{\rho }^2 \right)
       \nonumber  \\
& &  + p^3\,\rho \, \left( -39 - 195\,\rho  + 99\,{\rho }^2 -
     4\,{\rho }^3 \right) \nonumber \\
& & + p^4\,\rho \, \left( 16 + 147\,\rho  -
     411\,{\rho }^2 + 43\,{\rho }^3 + 11\,{\rho }^4
     \right) \nonumber \\
& &  -
  p^5\,\rho^2 \,\left( -54 - 2300\,\rho  +
     123\,{\rho }^2 + 67\,{\rho }^3 + 7\,{\rho }^4
     \right) \nonumber \\
& & -
  p^6\,\rho^2 \,\left( 126 + 4700\,\rho  +
     1865\,{\rho }^2 + 559\,{\rho }^3 -
     59\,{\rho }^4 + 13\,{\rho }^5 \right) \nonumber \\
& &   +
p^7\,\rho^2 \,
   \left( 41 + 4859\,\rho  + 2324\,{\rho }^2 +
     6400\,{\rho }^3 + 1135\,{\rho }^4 +
     189\,{\rho }^5 + 14\,{\rho }^6 \right) \nonumber \\
& &   +
  p^8\,{\rho }^3\,\left( -2701 + 4149\,\rho  -
     11762\,{\rho }^2 - 8744\,{\rho }^3 -
     2515\,{\rho }^4 - 311\,{\rho }^5 + 4\,{\rho }^6
     \right) \nonumber \\
& &     -
  p^9\,{\rho }^3\,\left( -781 + 12546\,\rho  -
     6054\,{\rho }^2 - 14564\,{\rho }^3 -
     11548\,{\rho }^4 - 3012\,{\rho }^5 -
     124\,{\rho }^6 + 8\,{\rho }^7 \right) \nonumber \\
& &  -
  p^{10}\,{\rho }^3\,\left( 96 - 13197\,\rho  -
     3996\,{\rho }^2 + 4727\,{\rho }^3 +
     17901\,{\rho }^4 + 11943\,{\rho }^5 +
     1464\,{\rho }^6 + 4\,{\rho }^7 \right) \nonumber \\
& &   +
  p^{11}\,{\rho }^4\,\left( -7185 - 3814\,\rho  -
     10626\,{\rho }^2 + 5695\,{\rho }^3 +
     21337\,{\rho }^4 + 5871\,{\rho }^5 +
     304\,{\rho }^6 \right) \nonumber \\
& &   -
  p^{12}\,{\rho }^4\,\left( -2040 + 3598\,\rho  -
     10478\,{\rho }^2 - 17573\,{\rho }^3 +
     19235\,{\rho }^4 + 11669\,{\rho }^5 +
     1219\,{\rho }^6 \right) \nonumber \\
& &   +
  p^{13}\,{\rho }^4\,\left( -240 + 6353\,\rho  +
     3091\,{\rho }^2 - 30176\,{\rho }^3 +
     6625\,{\rho }^4 + 14726\,{\rho }^5 +
     2128\,{\rho }^6 \right) \nonumber \\
& &  -
  p^{14}\,{\rho }^5\,\left( 3613 + 12361\,\rho  -
     25852\,{\rho }^2 - 5221\,{\rho }^3 +
     14188\,{\rho }^4 + 2147\,{\rho }^5 \right) \nonumber \\
& &  +
  p^{15}\,{\rho }^5\,\left( 947 + 11148\,\rho  -
     14935\,{\rho }^2 - 8341\,{\rho }^3 +
     9720\,{\rho }^4 + 2095\,{\rho }^5 \right) \nonumber \\
& &    -
  p^{16}\,{\rho }^5\,\left( 96 + 5829\,\rho  -
     7507\,{\rho }^2 - 1909\,{\rho }^3 +
     690\,{\rho }^4 + 3066\,{\rho }^5 \right) \nonumber \\
& & + 2\,
p^{17}\,{\rho }^6\,
   \left( 993 - 2513\,\rho  + 3568\,{\rho }^2 -
     4056\,{\rho }^3 + 2046\,{\rho }^4 \right) \nonumber \\
& &   -
  p^{18}\,{\rho }^6\,\left( 418 - 3991\,\rho  +
     10333\,{\rho }^2 - 10532\,{\rho }^3 +
     3785\,{\rho }^4 \right) \nonumber \\
& & + p^{19}\,{\rho }^6\,\left( 41 - 2397\,\rho  +
     7184\,{\rho }^2 - 7257\,{\rho }^3 +
     2430\,{\rho }^4 \right) \nonumber \\
& & + 3\,p^{20}\,\left( 1 - \rho  \right) \,{\rho }^7\,
   \left( 299 - 684\,\rho  + 367\,{\rho }^2 \right) \nonumber \\
& &  - p^{21}\,\left( 1 - \rho  \right) \,{\rho }^7\,
   \left( 185 - 519\,\rho  + 322\,{\rho }^2 \right) \nonumber \\
& &  + p^{22}\,\left( 1 - \rho  \right) \,{\rho }^7\,
   \left( 1 - 2\,\rho  \right) \,
   \left( 16 - 15\,\rho  \right) \nonumber \\
& & - 19\,p^{23}\,{\left( 1 - \rho  \right) }^2\,
   {\rho }^8 + 8\,p^{24}\, {\left( 1 - \rho  \right) }^2\,{\rho }^8 -
  p^{25}\,{\left( 1 - \rho  \right) }^2\,{\rho }^8
\label{polynome critique P1}
\end{eqnarray}
\normalsize

\subsection{${\mathcal P}_2(p,\rho)$}
\label{plimite2}

\small
\begin{eqnarray}
{\mathcal P}_2(p,\rho) & = & \nonumber \\
& & \nonumber \\
& & \nonumber \hskip-2cm
9  + 2\,p\,
   \left( -93 + 37\,\rho  \right) +
  p^2\,\left( 1375 - 1139\,\rho  + 235\,{\rho }^2
     \right) \nonumber \\
& & \hskip-2cm
 + p^3\,\left( -4548 + 6619\,\rho  -
     2306\,{\rho }^2 + 380\,{\rho }^3 \right) \nonumber \\
& & \hskip-2cm
   + p^4\,\left( 7551 - 22762\,\rho  +
     7011\,{\rho }^2 - 2415\,{\rho }^3 + 335\,{\rho }^4
     \right) \nonumber \\
& & \hskip-2cm
 +
  p^5\,\left( -6210 + 53479\,\rho  + 2160\,{\rho }^2 +
     7102\,{\rho }^3 - 1462\,{\rho }^4 + 154\,{\rho }^5
     \right) \nonumber \\
& & \hskip-2cm
   +
  p^6\,\left( 2025 - 85209\,\rho  - 79460\,{\rho }^2 -
     18121\,{\rho }^3 - 11141\,{\rho }^4 +
     223\,{\rho }^5 + 29\,{\rho }^6 \right) \nonumber \\
& & \hskip-2cm
+ p^7\,\rho \,
   \left( 84879 + 277027\,\rho  + 106511\,{\rho }^2 +
     202610\,{\rho }^3 - 36581\,{\rho }^4 +
     1170\,{\rho }^5 \right) \nonumber \\
& & \hskip-2cm
 +
  p^8\,\rho \,\left( -46170 - 521894\,\rho  -
     587852\,{\rho }^2 - 1536749\,{\rho }^3 +
     263741\,{\rho }^4 - 39391\,{\rho }^5 +
     547\,{\rho }^6 \right) \nonumber \\
& & \hskip-2cm
 -
  p^9\,\rho \,\left( -10125 - 611502\,\rho  -
     1935829\,{\rho }^2 - 7279427\,{\rho }^3 +
     440373\,{\rho }^4 \right. \nonumber \\
& & \hskip-0.5cm \left. - 331984\,{\rho }^5 +
     16780\,{\rho }^6 \right) \nonumber \\
& & \hskip-2cm
 +
  p^{10}\,{\rho }^2\,\left( -455796 - 3964687\,\rho  -
     23412205\,{\rho }^2 - 3663927\,{\rho }^3 -
     1452867\,{\rho }^4 \right. \nonumber \\
& & \hskip-0.5cm \left. + 375501\,{\rho }^5 +
     2506\,{\rho }^6 \right) \nonumber \\
& & \hskip-2cm
 +
  p^{11}\,{\rho }^2\,\left( 204255 + 5324974\,\rho  +
     53128592\,{\rho }^2 + 28420289\,{\rho }^3 +
     5054943\,{\rho }^4 \right. \nonumber \\
& & \hskip-0.5cm \left. - 4185113\,{\rho }^5 +
     12850\,{\rho }^6 \right) \nonumber \\
& & \hskip-2cm
 + p^{12}\,{\rho }^2\,
   \left( -42525 - 4776213\,\rho  -
     87000246\,{\rho }^2 - 104194583\,{\rho }^3 \right. \nonumber \\
& & \hskip-0.5cm \left. -
     21210055\,{\rho }^4 + 27622695\,{\rho }^5  -
     279589\,{\rho }^6 + 23497\,{\rho }^7 \right) \nonumber \\
& & \hskip-2cm
 +
  p^{13}\,{\rho }^3\,\left( 2798298 +
     103839446\,\rho  + 248730949\,{\rho }^2 +
     91081790\,{\rho }^3 \right. \nonumber \\
& & \hskip-0.5cm \left. - 119321953\,{\rho }^4 -
     214483\,{\rho }^5 + 2402\,{\rho }^6 \right) \nonumber \\
& & \hskip-2cm
 +
  p^{14}\,{\rho }^3\,\left( -987660 -
     90012828\,\rho  - 421343630\,{\rho }^2 -
     300983008\,{\rho }^3 \right. \nonumber \\
& & \hskip-0.5cm \left. + 354546288\,{\rho }^4 +
     19055623\,{\rho }^5 - 2902607\,{\rho }^6 +
     156356\,{\rho }^7 \right) \nonumber \\
& & \hskip-2cm
 -
  p^{15}\,{\rho }^3\,\left( -164025 -
     55522953\,\rho  - 523811627\,{\rho }^2 -
     718737692\,{\rho }^3 \right. \nonumber \\
& & \hskip-0.5cm \left. + 737559653\,{\rho }^4 +
     137144016\,{\rho }^5 - 24588235\,{\rho }^6 +
     1508228\,{\rho }^7 \right) \nonumber \\
& & \hskip-2cm
 +
  p^{16}\,{\rho }^4\,\left( -23264478 -
     483102731\,\rho  - 1256125498\,{\rho }^2 +
     1050744677\,{\rho }^3 \right. \nonumber \\
& & \hskip-0.5cm \left. + 534128347\,{\rho }^4 -
     99647028\,{\rho }^5 + 3934654\,{\rho }^6 +
     455372\,{\rho }^7 \right) \nonumber \\
& & \hskip-2cm
  -
  p^{17}\,{\rho }^4\,\left( -5993055 -
     328891636\,\rho  - 1637004249\,{\rho }^2 +
     897069980\,{\rho }^3 \right. \nonumber \\
& & \hskip-0.5cm \left. + 1365985483\,{\rho }^4 -
     226306853\,{\rho }^5 - 13885678\,{\rho }^6 +
     5675792\,{\rho }^7 \right) \nonumber \\
& & \hskip-2cm
 +
  p^{18}\,{\rho }^4\,\left( -729000 -
     161715291\,\rho  - 1608420692\,{\rho }^2 +
     82688597\,{\rho }^3 + 2449767599\,{\rho }^4 \right. \nonumber \\
& & \hskip-0.5cm \left. -
     232309325\,{\rho }^5 - 139531559\,{\rho }^6 +
     32366472\,{\rho }^7 + 274104\,{\rho }^8 \right) \nonumber \\
& & \hskip-2cm
-
  p^{19}\,{\rho }^5\,\left( -54779328 -
     1192868328\,\rho  - 1036350248\,{\rho }^2 +
     3126514952\,{\rho }^3 \right. \nonumber \\
& & \hskip-0.5cm \left. + 252124085\,{\rho }^4 -
     530576442\,{\rho }^5 + 110376920\,{\rho }^6 +
     3523400\,{\rho }^7 \right) \nonumber \\
& & \hskip-2cm
  +
  p^{20}\,{\rho }^5\,\left( -11563290 -
     661113157\,\rho  - 1778966082\,{\rho }^2 +
     2728489572\,{\rho }^3 \right. \nonumber \\
& & \hskip-0.5cm \left. + 1464568704\,{\rho }^4 -
     1261931324\,{\rho }^5 + 243789711\,{\rho }^6 +
     21533484\,{\rho }^7 \right) \nonumber \\
& & \hskip-2cm
 -
  p^{21}\,{\rho }^5\,\left( -1166400 -
     267060831\,\rho  - 1769360761\,{\rho }^2 +
     1310882671\,{\rho }^3 \right. \nonumber \\
& & \hskip-0.5cm \left. + 3038379428\,{\rho }^4 -
     2084661610\,{\rho }^5 + 338855326\,{\rho }^6 +
     82890944\,{\rho }^7 \right) \nonumber \\
& & \hskip-2cm
  +
  p^{22}\,{\rho }^6\,\left( -74824812 -
     1221905422\,\rho  - 276377214\,{\rho }^2 +
     4066473373\,{\rho }^3 \right. \nonumber \\
& & \hskip-0.5cm \left. - 2443698651\,{\rho }^4 +
     202709670\,{\rho }^5 + 224730742\,{\rho }^6
     \right) \nonumber \\
& & \hskip-2cm
 - p^{23}\,{\rho }^6\,
   \left( -13126860 - 609332834\,\rho  -
     1166508009\,{\rho }^2 + 3871377708\,{\rho }^3 \right. \nonumber \\
& & \hskip-0.5cm \left. -
     1932786432\,{\rho }^4 - 295096153\,{\rho }^5 +
     454945390\,{\rho }^6 \right) \nonumber \\
& & \hskip-2cm
 +
  p^{24}\,{\rho }^6\,\left( -1103625 -
     218082828\,\rho  - 1184923715\,{\rho }^2 +
     2650067734\,{\rho }^3 \right. \nonumber \\
& & \hskip-0.5cm \left. - 752045795\,{\rho }^4 -
     1022894704\,{\rho }^5 + 711726527\,{\rho }^6
     \right)  \nonumber \\
& & \hskip-2cm
  -
  p^{25}\,{\rho }^7\,\left( -53743284 -
     754572613\,\rho  + 1226487190\,{\rho }^2 +
     416790514\,{\rho }^3 \right. \nonumber \\
& & \hskip-0.5cm \left. - 1614449937\,{\rho }^4 +
     878716880\,{\rho }^5 \right) \nonumber \\
& & \hskip-2cm
 +
  p^{26}\,{\rho }^7\,\left( -8258625 -
     336506619\,\rho  + 274671788\,{\rho }^2 +
     1010093115\,{\rho }^3 \right. \nonumber \\
& & \hskip-0.5cm \left. - 1758220080\,{\rho }^4 +
     866894232\,{\rho }^5 \right) \nonumber \\
& & \hskip-2cm
 -
  p^{27}\,{\rho }^7\,\left( -605475 -
     106456428\,\rho  - 91817943\,{\rho }^2 +
     978393477\,{\rho }^3 \right. \nonumber \\
& & \hskip-0.5cm \left. - 1446294368\,{\rho }^4 +
     687560517\,{\rho }^5 \right) \nonumber \\
& & \hskip-2cm
  + p^{28}\,{\rho }^8\,
   \left( -23082921 - 120749456\,\rho  +
     640575958\,{\rho }^2 - 928058750\,{\rho }^3 +
     438708098\,{\rho }^4 \right) \nonumber \\
& & \hskip-2cm
 -
  p^{29}\,{\rho }^8\,\left( -3116070 -
     61389864\,\rho  + 311935039\,{\rho }^2 -
     469461042\,{\rho }^3 + 224128938\,{\rho }^4
     \right) \nonumber \\
& & \hskip-2cm
+ p^{30}\,{\rho }^8\,
   \left( -200475 - 19494900\,\rho  +
     116248678\,{\rho }^2 - 186770845\,{\rho }^3 +
     90668832\,{\rho }^4 \right) \nonumber \\
& & \hskip-2cm
   - p^{31}\,{\rho }^9\,
   \left( -4042170 + 33260499\,\rho  -
     57626524\,{\rho }^2 + 28477028\,{\rho }^3 \right) \nonumber \\
& & \hskip-2cm
 +
  p^{32}\,{\rho }^9\,\left( -510300 + 7194033\,\rho  -
     13394778\,{\rho }^2 + 6717655\,{\rho }^3 \right) \nonumber \\
& & \hskip-2cm
-
  3\,p^{33}\,{\rho }^9\,
   \left( -10125 + 376209\,\rho  - 740598\,{\rho }^2 +
     374614\,{\rho }^3 \right) \nonumber \\
& & \hskip-2cm
 +
  540\,p^{34}\,\left( 1 - \rho  \right) \,
   {\rho }^{10}\,\left( 216 - 221\,\rho  \right)
- 6075\,p^{35}\,{\left( 1 - \rho  \right) }^2\,
   {\rho }^{10}.
\label{polynome critique P2}
\end{eqnarray}
\normalsize

\subsection{${\mathcal P}_3(p, \rho, T)$}
\label{plimite3}

\small
\begin{eqnarray}
{\mathcal P}_3(p, \rho, T) & = & \nonumber \\
& & \nonumber \\
& & \hskip-2cm 2 + 2\,T + p\,\left( -6 - 10\,T - 4\,T^2 \right) \nonumber \\
& & \hskip-2cm +
  p \,\rho\,\Big[ 5 + 12\,T - 8\,T^3 +
     p\,\left( -29 - 52\,T + 24\,T^2 + 64\,T^3 +
        16\,T^4 \right) \nonumber \\
& & \hskip-0.9cm  +
     p^2\,\left( 78 + 138\,T - 52\,T^2 - 176\,T^3 -
        64\,T^4 \right)  \nonumber \\
& & \hskip-0.9cm  +
     p^3\,\left( -36 - 60\,T + 48\,T^2 + 120\,T^3 +
        48\,T^4 \right) \Big]  \nonumber \\
& & \hskip-2cm +
  p^2 \,{\rho }^2\,\Big[ 4\,T + 8\,T^2 - 8\,T^3 - 16\,T^4 +
     p\,\left( -8 + 8\,T - 24\,T^2 + 48\,T^4 \right)
         \nonumber \\
& & \hskip-0.5cm + p^2\,\left( 56 - 22\,T + 44\,T^2 +
        192\,T^3 + 128\,T^4 + 64\,T^5 \right)  \nonumber \\
& & \hskip-0.5cm  +
     p^3\,\left( -262 - 336\,T + 40\,T^3 - 256\,T^4 -
        288\,T^5 - 64\,T^6 \right) \nonumber \\
& & \hskip-0.5cm  +
     p^4\,\left( 248 + 384\,T - 120\,T^2 - 384\,T^3 +
        128\,T^4 + 384\,T^5 + 128\,T^6 \right)  \nonumber \\
& & \hskip-0.5cm  +
     p^5\,\left( -66 - 110\,T + 52\,T^2 + 160\,T^3 -
        32\,T^4 - 160\,T^5 - 64\,T^6 \right)  \Big]
      \nonumber \\
& & \hskip-2cm + p^3 \,{\rho }^3\,
  \Big[ -10 - 40\,T - 40\,T^2 + 8\,T^3 + 16\,T^4  \nonumber \\
& & \hskip-0.5cm  +
     p\,\left( 58 + 164\,T + 264\,T^2 + 24\,T^3 -
        256\,T^4 - 128\,T^5 \right)  \nonumber \\
& & \hskip-0.5cm  +
     p^2\,\left( -206 - 378\,T - 800\,T^2 - 656\,T^3 +
        416\,T^4 + 640\,T^5 + 192\,T^6 \right)   \nonumber \\
& & \hskip-0.5cm  +  p^3\,\left( 448 + 890\,T + 1296\,T^2 +
        1080\,T^3 - 256\,T^4 - 864\,T^5 - 384\,T^6
        \right)
  \nonumber \\
& & \hskip-0.5cm  + p^4\,
      \left( -237 - 668\,T - 1188\,T^2 - 920\,T^3 -
        32\,T^4 + 320\,T^5 + 192\,T^6 \right) \nonumber  \\
& & \hskip-0.5cm
 +
     p^5\,\left( -151 - 70\,T + 760\,T^2 + 904\,T^3 +
        256\,T^4 + 32\,T^5 \right)   \nonumber \\
& & \hskip-0.5cm  +
     p^6\,\left( 162 + 230\,T - 300\,T^2 - 560\,T^3 -
        192\,T^4 \right)   \nonumber \\
& & \hskip-0.5cm +
     p^7\,\left( -36 - 60\,T + 48\,T^2 + 120\,T^3 +
        48\,T^4 \right)   \Big]
      \nonumber \\
& & \hskip-2cm + p^4 \,{\rho }^4\,
   \Big[ -10 - 54\,T - 104\,T^2 - 56\,T^3 + 80\,T^4 +
     64\,T^5   \nonumber \\
& & \hskip-0.5cm  +
     p\,\left( 54 + 274\,T + 492\,T^2 + 400\,T^3 -
        144\,T^4 - 416\,T^5 - 192\,T^6 \right) \nonumber  \\
& & \hskip-0.5cm  + p^2\,
      \left( -12 - 578\,T - 1244\,T^2 - 904\,T^3 +
        48\,T^4 + 576\,T^5 + 384\,T^6 \right)
     \nonumber \\
& & \hskip-0.5cm  +
     p^3\,\left( -206 + 456\,T + 2000\,T^2 +
        1688\,T^3 + 304\,T^4 - 160\,T^5 - 192\,T^6
        \right) \nonumber  \\
& & \hskip-0.5cm  +
     p^4\,\left( 156 - 356\,T - 2060\,T^2 -
        2120\,T^3 - 592\,T^4 - 64\,T^5 \right) \nonumber  \\
& & \hskip-0.5cm  +
     p^5\,\left( 92 + 462\,T + 1228\,T^2 + 1272\,T^3 +
        400\,T^4 \right)   \nonumber \\
& & \hskip-0.5cm + p^6\,\left( -88 - 270\,T - 372\,T^2 -
        288\,T^3 - 96\,T^4 \right)  \nonumber \\
& & \hskip-0.5cm   +
     p^7\,\left( -22 - 2\,T + 28\,T^2 + 8\,T^3 \right) \nonumber  \\
& & \hskip-0.5cm + p^8\,\left( 30 + 46\,T + 16\,T^2 \right)  \nonumber \\
& & \hskip-0.5cm  + p^9\,\left( -6 - 10\,T - 4\,T^2 \right) \Big] \nonumber  \\
& & \hskip-2cm + p^5 \,{\rho }^5\,
   \Big[ -3 - 20\,T - 56\,T^2 - 64\,T^3 - 16\,T^4 +
     64\,T^5 + 64\,T^6 \nonumber \\
& & \hskip-0.5cm  +
     p\,\left( 11 + 160\,T + 432\,T^2 + 376\,T^3 +
        80\,T^4 - 96\,T^5 - 128\,T^6 \right) \nonumber \\
& & \hskip-0.5cm   +
     p^2\,\left( -84 - 488\,T - 1116\,T^2 - 984\,T^3 -
        240\,T^4 + 64\,T^6 \right) \nonumber  \\
& & \hskip-0.5cm   +
     p^3\,\left( 240 + 850\,T + 1416\,T^2 +
        1224\,T^3 + 336\,T^4 + 32\,T^5 \right) \nonumber  \\
& & \hskip-0.5cm   + p^4\,\left( -252 - 790\,T - 960\,T^2 -
        712\,T^3 - 208\,T^4 \right) \nonumber  \\
& & \hskip-0.5cm  +
     p^5\,\left( 74 + 318\,T + 328\,T^2 + 168\,T^3 +
        48\,T^4 \right)  \nonumber \\
& & \hskip-0.5cm  +
     p^6\,\left( 43 + 14\,T - 28\,T^2 - 8\,T^3 \right) \nonumber \\
& & \hskip-0.5cm  +
     p^7\,\left( -33 - 48\,T - 16\,T^2 \right)  \nonumber \\
& & \hskip-0.5cm  +
     p^8\,\left( 6 + 10\,T + 4\,T^2 \right) \Big]
\label{polynome critique P3}
\end{eqnarray}
\normalsize

When $T = -1$, we can further simplify the above expression 
\begin{equation}
{\mathcal P}_3(p,\rho,T = -1) = \left( 1 - p \right) \,p\,\rho \,
  {\Big( 1 + p\,(1-p) \, \rho \Big) }^2\,
  {\Big( 1 - \rho + \rho \, (1-p)^3 \Big) }^2
\label{P3 avec T = -1}
\end{equation}

There is no such factorization in the case $T=1$:
\small
\begin{eqnarray}
{\mathcal P}_3(p,\rho,T = 1) & = &  \nonumber \\
& &  \nonumber \\
& & \hskip-3cm 4 - 20\,p + p\, \rho  \, \left( 9 + 23\,p - 76\,p^2 + 120\,p^3 \right) \nonumber \\
& & \hskip-3cm  - 2\,p^2\, {\rho }^2 \,\left( 6 - 12\,p - 231\,p^2 + 583\,p^3 - 384\,p^4 + 110\,p^5 \right) \nonumber \\
& & \hskip-3cm + p^3 \,{\rho }^3 \,\left( -66 + 126\,p - 792\,p^2 + 2210\,p^3 - 2533\,p^4 + 1731\,p^5 - 660\,p^6 + 120\,p^7 \right) \nonumber  \\
& & \hskip-3cm - 2\, p^4 \, {\rho }^4\,\left( 40 - 234\,p + 865\,p^2 - 1945\,p^3 + 2518\,p^4 - 1727\,p^5 + 557\,p^6 - 6\,p^7 - 46\,p^8 + 10\,p^9 \right) \nonumber \\
& & \hskip-3cm + p^5 \, {\rho }^5 \, \left( -31 + 835\,p - 2848\,p^2 + 4098\,p^3 - 2922\,p^4 + 936\,p^5 + 21\,p^6 - 97\,p^7 + 20\,p^8 \right)
\label{polynome limite 1}
\end{eqnarray}
\normalsize


\end{document}